\documentclass[draft]{agujournal2019}

\newcommand{\aap}{Astron. Astrophys.}

\usepackage{url} %this package should fix any errors with URLs in refs.
\usepackage{lineno}
\usepackage[inline]{trackchanges} %for better track changes. finalnew option will compile document with changes incorporated. Ki
\usepackage{ragged2e}
\usepackage{soul}
\usepackage{amsthm}
\usepackage{amssymb}
\usepackage{amsmath}
\usepackage{bm}

\draftfalse

\journalname{JGR: Space Physics}

\justifying
\begin{document}

%%%%%%%%%%%%%%%%%%%%%%%%%%%%%%%%%%%%%%%%%%%%%%%
%  TITLE
%%%%%%%%%%%%%%%%%%%%%%%%%%%%%%%%%%%%%%%%%%%%%%%
\title{Extended scenarios for solar radio emissions with downshifted electron beam plasma excitations}

%%%%%%%%%%%%%%%%%%%%%%%%%%%%%%%%%%%%%%%%%%%%%%%
%  AUTHORS AND AFFILIATIONS
%%%%%%%%%%%%%%%%%%%%%%%%%%%%%%%%%%%%%%%%%%%%%%%

\authors{M. Lazar\affil{1,2}, 
         R.~A. L\'{o}pez\affil{3,4},
         S.M. Shaaban\affil{5},
         S.Poedts\affil{1,6},
         H. Fichtner\affil{7}
          }
\affiliation{1}{Centre for Mathematical Plasma Astrophysics, KU Leuven, Celestijnenlaan 200B, B-3001 Leuven, Belgium}
\affiliation{2}{Institut f\"ur Theoretische Physik IV, Ruhr-Universit\"at Bochum, D-44780 Bochum, Germany}
\affiliation{3}{ Research Center in the intersection of Plasma Physics, Matter, and Complexity ($P^2 mc$), Comisi\'on Chilena de Energ\'{\i}a Nuclear, Casilla 188-D, Santiago, Chile}
\affiliation{4}{Departamento de Ciencias F\'{\i}sicas, Facultad de Ciencias Exactas, Universidad Andres Bello, Sazi\'e 2212, Santiago 8370136, Chile}
\affiliation{5}{Department of Physics and Materials Sciences, College of Arts and Sciences, Qatar University, 2713 Doha, Qatar}
\affiliation{6}{Institute of Physics, University of Maria Curie-Sk{\l}odowska, ul.\ Marii Curie-Sk{\l}odowskiej 1, 20-031 Lublin, Poland}
\affiliation{7}{Institut f\"ur Theoretische Physik IV, Ruhr-Universit\"at Bochum, D-44780 Bochum, Germany}

%(repeat as many times as is necessary)

% Corresponding author mailing address and e-mail address:

% (include name and email addresses of the corresponding author.  More
% than one corresponding author is allowed in this LaTeX file and for
% publication; but only one corresponding author is allowed in our
% editorial system.)

\correspondingauthor{S.M.Shaaban}{shamd@qu.edu.qa; s.m.shaaban88@gmail.com}

%%%%%%%%%%%%%%%%%%%%%%%%%%%%%%%%%%%%%%%%%%%%%%%
% KEY POINTS
%%%%%%%%%%%%%%%%%%%%%%%%%%%%%%%%%%%%%%%%%%%%%%%
%  List up to three key points (at least one is required)

\begin{keypoints}
\item Denser or colder electron beams than in the standard plasma emission model generate downshifted excitations, observed in terrestrial foreshock.
%Electron beams denser or colder than in the standard plasma emission model generate downshifted primary excitations, observed already in terrestrial foreshock.
\item Kinetic theory and simulations suggest direct and indirect involvement of downshifted excitations in the generation of radio emissions.
\item The downshifted excitations are involved directly in the generation of the second radio harmonic, very effectively for systems with two electron counter-beams.
\end{keypoints}

%%%%%%%%%%%%%%%%%%%%%%%%%%%%%%%%%%%%%%%%%%%%%%%
%
%  ABSTRACT and PLAIN LANGUAGE SUMMARY

% The Plain Language Summary should be written for a broad audience,
% including journalists and the science-interested public, that will not have 
% a background in your field.
%
% A Plain Language Summary is required in GRL, JGR: Planets, JGR: Biogeosciences,
% JGR: Oceans, G-Cubed, Reviews of Geophysics, and JAMES.
% see http://sharingscience.agu.org/creating-plain-language-summary/)
%
%%%%%%%%%%%%%%%%%%%%%%%%%%%%%%%%%%%%%%%%%%%%%%%

%% \begin{abstract} starts the second page

\begin{abstract}
First-principle studies of radiative processes aimed at explaining the origin of type II and type III solar radio bursts raise questions on the implications of downshifted electron beam plasma excitations with frequency (slightly) below the plasma frequency ($\omega\lesssim\omega_{pe}$) in the generation of radio emissions. 
Unlike the beam-induced Langmuir waves ($\omega \gtrsim \omega_{pe}$) in the standard radio emission plasma model, the primary wave excitations of cooler and/or denser beams have predominantly downshifted frequencies.
Broadbands of such downshifted excitations are also confirmed by in situ observations in association with terrestrial foreshock and electron beams (in contrast to narrowband Langmuir waves), but their involvement in radiative processes has not been examined so far.
We revisit three radiative scenarios specific to downshifted primary excitations, and the results demonstrate their direct or indirect involvement in plasma radio emission. 
Downshifted excitations of an electron beam primarily play an indirect role, contributing to the relaxation to a plateau-on-tail still able to induce Langmuir beam waves that satisfy conditions for nonlinear wave-wave interactions leading to free radio waves.
At longer time scales, the primary excitations can become predominantly downshifted, and then directly couple with the secondary (backscattered) Langmuir waves to generate the second harmonic of radio emissions.
Two counterbeams are more efficient and lead to faster radiative mechanisms, involving counterpropagating downshifted excitations, which couple to each other and generate intense, broadband and isotropic radio spectra of downshifted second harmonics.
Such a long-lasting (second) radio harmonic can thus be invoked to distinguish regimes with downshifted ($\omega \gtrsim \omega_{pe}$) primary excitations.
%Such in-depth analyzes unveil the complexity of the radiative processes by the electron beam-plasma interaction, with the unexpected involvement of downshifted excitations.
%The results suggest an extension of the standard plasma model applied to solar radio emissions to include new regimes involving downshifted excitations of solar electron beams.
%These models assume less extreme conditions, i.e. electron beams/strahls with moderate number densities and beam velocities similar to those associated with sources of type II and type III solar bursts. The ES high-frequency excitations have a predominantly downshifted frequency slightly below $\omega_{pe}$, and belong to the branch of beam modes with higher growth rates than modified-Langmuir waves. Moreover, the implication of low-frequency waves as ion-acoustic waves remains inconclusive, while numerical simulations suggest that radio emissions can be directly generated by linear or nonlinear wave-wave coupling without any parametric decay process.
\end{abstract}

\section*{Plain Language Summary}
In cosmology, astrophysics, but also in the solar and geophysical contexts, radio emissions are true messengers of their distant sources, whose interpretation is generally based on the standard radio emission model of the electron-beam plasmas exciting Langmuir electrostatic waves. With frequencies higher than the plasma frequency, intense Langmuir waves can directly engage in wave-wave interactions, leading to escaping radio electromagnetic waves. Present analysis strongly suggests that primary excitations with lower frequencies but still close to the plasma frequency may also contribute, directly or indirectly, to the generation of radio emissions. These downshifted excitations are confirmed by the observations, but are produced by denser and/or cooler electron beams than in the standard model, markedly expanding the parametric regimes to be considered in the remote diagnosis of radio sources.
%Here are instructions on writing a Plain Language Summary: https://www.agu.org/Share-and-Advocate/Share/Community/Plain-language-summary

%%%%%%%%%%%%%%%%%%%%%%%%%%%%%%%%%%%%%%%%%%%%%%%
%  BODY TEXT
%%%%%%%%%%%%%%%%%%%%%%%%%%%%%%%%%%%%%%%%%%%%%%
%%%
\section{Motivation} \label{sec-1}
%%%
%
Radio emissions represent a topic of great interest in space and astrophysical applications, especially for the exploration of plasma sources that are inaccessible to in situ observations \cite{Warmuth-2005, Cremades-2015, Crosley_2016, Mann-eta-2018, Villadsen_2019, Davis-etal-2021}. 
However, indirect remote diagnostics requires an understanding of the physical mechanisms that produce radio emissions and, implicitly, realistic modeling of plasma systems and the physics involved \cite{Nindos-2008, Pick-2008, Reid-2014}. 
The most cited source of radio emissions are the electron beams released by energetic solar events, relevant being type II radio bursts associated with interplanetary shocks generated by coronal mass ejections (CMEs), but also type III radio bursts triggered by energetic electrons from coronal flares \cite{Lin-etal-1981, Bale-etal-1999, Pulupa-Bale-2008, Mann-eta-2018}.
The ability of electron beams to produce radio emissions is predicted by theory \cite{Cairns-1989, Ziebel-etal-2014, Ziebell-etal-2016, Lazar-etal-2023a}, and is ultimately proven by numerical simulations \cite{Kasaba-etal-2001, Rhee-etal-2009, Umeda-2010, Ganse-etal-2012a, Ganse-etal-2012b, Thurgood-Tsiklauri-2015, Henri-etal-2019, Sauer-etal-2019, Krafft_2021, Krafft_2022, Lee-etal-2022, Lazar-etal-2023a, Bachini-2024}.
The linear (kinetic) theory identifies the primary wave excitations of electron beams \cite{Cairns-1989, Lazar-etal-2023a}, while, for example, a weak turbulence (WT) approach explains their nonlinear wave-wave conversion into radio electromagnetic (EM) waves \cite{Ziebel-etal-2014, Ziebell-etal-2016, Lee-etal-2019}.
Theory is therefore crucial for the interpretation of PIC simulations as well as the observations.
The most invoked are plasma emission mechanisms centered on primary excitations of Langmuir waves (symbolized by $L$ hereafter)  with frequencies near and above the plasma frequency, $\omega \gtrsim \omega_{pe}$ \cite{Melrose_2008}.
At high intensities, $L$ waves satisfy wave-wave decays, leading to daughter waves, i.e., ion-sound ($S$) waves and secondary $L'$ waves, $L \to S + L'$, which will then allow couplings (coalescence) to generate EM or transverse ($T$) waves via, e.g., $L+S \to T = F$ for the fundamental emission ($F$), and $L+L' \to 2T = H$ for the second order harmonic ($H$). 

The primary excitations are, however, highly dependent on the properties of the electron beams, such as their density, bulk velocity or drift relative to the bulk or core population, thermal and suprathermal spread, and, nevertheless, the beam configuration, as either a singular beam aligned to an open magnetic field or two counter-beams (bi-directional electrons) in, e.g., a closed magnetic field topology.
On the one hand, the excitation of $L$ waves is limited to a narrow parametric regime, i.e., electron beams (subscript $b$) with very low number densities $n_b$ and whose speeds $U_b$ satisfy \cite{Cairns-1989, Gary-1993}
\begin{align}
 \theta_b < U_b < (n_e/n_b)^{1/3} \theta_b, \label{e1}  
\end{align}
where $\theta_b = (2\pi k_B T_b/m_e)^{1/2}$ is the thermal spread (or velocity) and $T_b$ the temperature of beam population, $m_e$ is the electron mass and $n_e$ the total number density of electrons.
On the other hand, for certain setups in PIC simulations of radio emissions, the electron beams trigger in the preliminary phase instabilities of electron beam (EB) modes, with downshifted frequencies slightly below the plasma frequency, $\omega \lesssim \omega_{pe}$ \cite{Sauer-etal-2019, Lazar-etal-2023a}.
In this case, we deal with cooler and/or denser beams, which generally satisfy 
\begin{align}
 U_b > (n_e/n_b)^{1/3} \theta_b, \label{e2}   
\end{align}
and excite yet resonant electrostatic (ES) waves from the branch of the electron-beam mode with (an almost) linear wavenumber ($k$) dispersion of the wave frequency ($\omega$), $\omega \simeq k U_b$.
Moreover, these setups still comply with observational estimations for the magnetized electron-beam plasma parameters in the source regions of solar radio bursts.
Therefore, such results raise questions about downshifted waves as primary excitations, if they are involved, and if so, in what processes they are involved in producing radio emissions. 
Because downshifted $EB$ waves contrast with the primary excitations in the standard model of plasma radio emission, which are generally assimilated to $L$-type waves with a sufficiently high frequency $\omega \gtrsim \omega_{pe}$.
Unlike $L$ waves, downshifted waves (with $\omega \lesssim \omega_{pe}$) cannot undergo wave-wave decays specific to standard plasma emission, failing to satisfy the Manley-Rowe laws of conservation for energy and momentum.

The ES plasma oscillations with downshifted frequencies are observed in the terrestrial foreshock or upstream regions in association with electron beams as broadband fluctuations that contrast strongly with narrowband $L$ waves \cite{Fuselier-etal-1985, Onsager-Holzworth-1990, Soucek-etal-2019}. 
To our knowledge, there are no similar in situ observations of interplanetary foreshocks, but the analyses of bursty radio emissions from the Earth foreshock \cite{Lacombe-etal-1988, Reiner-etal-1997, Kasaba-etal-2000} in conjunction with in situ observations strongly suggest that the underlying physics of electron beams and induced waves is common to all planetary and interplanetary foreshocks, including type II radio burst sources \cite{Boshuizen-etal-2004, Kuncic-etal-2005}. 
Moreover, the parameters of electron beams associated with downshifted ES excitations \cite{Soucek-etal-2019} appear to be similar to the sources of type II radio emission \cite{Pulupa-Bale-2008}, as also noted by \cite{Lazar-etal-2023a}.

Early theories and numerical simulations also support generation of downshifted waves, although the (initial) beams are often assumed to be much cooler than the core population, with, i.e., peak-on-tail rather than bump-on-tail distributions (presumably closer to the source of beam injection than the more relaxed states observed in situ)  \cite{Fuselier-etal-1985, Onsager-Holzworth-1990, Dum-1990a, Dum-1990b}.
The simulations generally confirm a transition from reactive or weakly resonant beam modes, with frequencies below plasma frequency $\omega_{pe}$, to the kinetic instability of Langmuir-beam ($LB$) waves with $\omega \sim \omega_{pe}$, corresponding to the beam relaxation to a more gentle bump-on-tail and even a flattened plateau-on-tail distribution \cite{Dum-1990a, Dum-1990b, Thurgood-Tsiklauri-2015, Sauer-etal-2019}.
Under the wave-particle interaction, the beam slows down and widens at the same time, and the velocity distributions depart significantly from the Maxwellian beam-plasma components assumed initially.
The kinetic regime seems to be more robust, lasting much longer than the broadband reactive regime, and downshifted excitations with frequency below but still near plasma frequency ($\omega \lesssim \omega_{pe}$) persist for a long time until the bump in the electron distribution is completely flattened \cite{Dum-1990a, Dum-1990b}.
These profiles resemble those from the evolution of gentle beams or marginally stable (plateaued) distributions, which are also associated with periods of local $L$ excitations during solar type III radio bursts \cite{Ergun_1998}.

\begin{table} [t]
\centering
\caption{Electron and proton plasma parameterization for the three cases investigated here.  Two electron configurations are distinguished: core-beam (C-B) in cases~1 and 2 and core-counterbeams (C-CBs) in case~3. The $EB$ instability is also characterized by the maximum growth rate ($\gamma_m/\omega_{pe}$) and the corresponding frequency ($\omega_m/\omega_{pe}$) and wave number ($\theta_e k_m/\omega_{pe}$) obtained from linear theory \cite{Lazar-etal-2023a}.} \label{t1}
\begin{tabular}{lccc}  
\hline
Parameters \,\,  $\backslash$   \,\, Cases &  1 (C-B) & 2 (C-B) & 3 (C-CBs)\\
   \hline
$T_e = T_c = T_{b}$  (10$^6$ K)& 2.32 & 2.32  & 2.32 \\
$\theta_e = \theta_c = \theta_{b}$ (10$^6$ m/s) & 8.4 & 8.4 & 8.4 \\
$T_p = 0.73 \, T_e$ (10$^6$ K) & 1.70 & 1.70  & 1.70 \\
$N_{b} = n_b /n_e$ & 0.0057 & 0.05  & 0.05 \\
$U_b/\theta_e$  & 16 & 8  & 8 \\
$\gamma_m/\omega_{pe}$  & 0.104 & 0.196  & 0.195 \\
$\omega_m/\omega_{pe}$  & 0.951 & 0.897  & 0.877 \\
$\theta k_m/\omega_{pe}$  & 0.066 & 0.142  & 0.139 \\
  \hline
\end{tabular}
\end{table}

For hot beams, with temperatures similar to that of the core population\footnote{Hotter or suprathermalized beams/strahls with $U_b \leqslant \theta$, are typical to the solar wind (in the absence of energetic events), and are rather susceptible to EM or hybrid excitations, known as firehose and whistler heat-flux instabilities \cite{Lopez-etal-2020}.}, the excitation of downshifted electron beam ($EB$) modes implies much higher relative drifts or beam speeds \cite{Sauer-Sydora-2012, Soucek-etal-2019, Sauer-etal-2019, Lopez-etal-2020, Lazar-etal-2023a, Lazar-etal-2023b}. 
Although primary excitations can also be $EB$ modes with (slightly) downshifted frequencies, $\omega \lesssim \omega_{pe}$, emphasis was placed on $LB$ excitations with frequencies $\omega \gtrsim \omega_{pe}$ triggered by the subsequent, more relaxed (plateaued) distribution \cite{Thurgood-Tsiklauri-2015, Sauer-etal-2019}. 
%This curve starts with the (almost) linear branch, $\omega \sim k U_b$, and after the shoulder (at $\sim \omega_{pe}$), it continues with the Langmuir-beam branch \cite{Cairns-1989}.
The later emerges from the standard dispersion curve for $L$ waves ($\omega^2 = \omega_{pe}^2 + 3 k^2 \theta_e^2$) in a plasma without a beam component \cite{Baumgartel-2014, Thurgood-Tsiklauri-2015}.
For sufficiently dense or/and cool beams, as our setups in Table~\ref{t1}, the $L$ dispersion curve in the (forward) direction of the beam (i.e., $k_\parallel >0$), breaks into two branches \cite{Cairns-1989}.
One is the $LB$ branch (also known as the electron-acoustic branch) at large $k > \omega_{pe}/U_b$, and the other in the optical domain (small $k$) couples to the linear dispersion of the $EB$ modes $\omega \simeq k (U_b + \theta_e)$, which increases steeply (above $\omega_{pe}$) with $k$.
The high frequency waves from the $LB$ branch are roughly described by $(\omega-kU_b)^2 \simeq \omega_{pe}^2 + k^2 \theta^2$ \cite{Swanson-2003}, and are not destabilized by the electron beam.
These two branches depart from each other for denser or/and cooler beams.
%Clearly, this branch is not significantly modified by the presence of the beam, but to distinguish in the case of branch brakes caused by denser or more energetic beams, it can fairly be called the Langmuir-beam branch \citep{Cairns-1989}, see also the explanations further below.
Only very weak beams, with very low densities, e.g., $n_b/n_e < 0.005$, can destabilize $L$ or $LB$ waves, because their dispersion curve ($\omega^2 \simeq \omega_{pe}^2 + 3 k^2 \theta_e^2$) does not break, but only undulates near unstable wavenumbers \cite{Cairns-1989, Lazar-etal-2023b}. 
In the backward direction ($k_\parallel< 0$), the $L$ mode is not affected by the beam, but is still distinguishable in the quasithermal fluctuations and the parametric (secondary) excitations.
At oblique propagation and large $k$, both the $LB$ (forward) and $L$ (backward) dispersion curves merge with the $Z$-mode, becoming what is called the Langmuir/Z ($LZ$) mode \cite{Benson-etal-2006}.
In contrast to downshifted $EB$ excitations, $L$ and $LB$ waves are, indeed, typical of the plasma emission model, which includes nonlinear decay $L \to S + L'$ into secondary $L'$ oscillations with low wavenumber $k_{L'} \to 0$, and $S$ waves with $k_S \sim k_L$. Further nonlinear couplings can produce transverse radio waves, both fundamental and harmonic emissions (see above). 
Moreover, long-lived (or free-damping) $L$ waves of low wavenumbers ($k_{L}c \sim \omega_{pe}$) can couple and linearly convert to EM radio waves with oblique propagation \cite{Sauer-etal-2019}. 
This process can last longer until the plateau relaxes, which provides a plausible explanation for Sturrock's dilemma \cite{sturrock-1964}.

In the present paper, we build on such results in, for instance, \citeA{Sauer-etal-2019} and \citeA{Lazar-etal-2023a}, seeking to clarify the role of downshifted plasma excitations in radio wave production.
We have refined and extended particle-in-cell (PIC) simulations for the three cases in \citeA{Lazar-etal-2023a}, to examine configurations with a single (asymmetric) electron beam and also two symmetric counterbeams (see Table~\ref{t1}), involving distinct radiative processes. 
Electron counterbeams typically produce radio harmonics on shorter time scales and with higher intensities \cite{Umeda-2010, Ganse-etal-2012b}, in this case, broadband second harmonics with downshifted frequencies, see also \citeA{Lazar-etal-2023a}.
%The radio emission of plasma systems with two counterbeams of electrons has been investigated in numerical simulations \cite{Ganse-etal-2012b, Thurgood-Tsiklauri-2015, Lazar-etal-2023a}, and also using weak turbulence theory \cite{Ziebell-etal-2016}.
%The second H emission reaches high intensities in the case of counter-propagating beams of equal properties, while the efficiency of F emission enhances when the lower density beam has a greater speed than the higher density beam \cite{Ziebell-etal-2016}.
However, none of the previous studies investigated the involvement of downshifted excitations, neither primary excitations nor radio EM waves.
Studies of radio emissions often assume similar or even identical parameterizations of electron beam plasmas, but mostly in the absence of an ambient magnetic field and without reference to downshifted excitations \cite{Baumgartel-2014, Thurgood-Tsiklauri-2015, Henri-etal-2019}.
The influence of the magnetic field ($B_0$) was also examined in PIC simulations, for strong fields or underdense plasmas with $\omega_{pe} \leqslant |\Omega_e|$, where $\Omega_e$
%= e B_0/(m_e c)$ 
is the electron gyrofrequency, and for weakly magnetized or overdense plasmas with $\omega_{pe} > |\Omega_e|$ \cite{Ganse-etal-2014, Sauer-etal-2019, Lee-etal-2022}. 
Radiative processes involving linear and nonlinear wave interactions remain the same in nature, and only the spectra of primary and radio wave excitations undergo changes.
Here, we provide a plausible explanation for the enhancement of the second radio harmonic in the presence of the magnetic field \cite{Lee-etal-2022}.

We use an explicit 2D PIC code based on the KEMPO1 code of \citeA{Matsumoto1993}. 
%
%It is thus suggested that the standard model of plasma emission should include cooler and/or denser electron beams at the origin of downshifted primary waves, thereby expanding the physical conditions conducive to radio emissions in the plasma sources of solar radio bursts.
%
The basic setup in the simulations consists of the following: a spatial domain composed of $1024\times1024$ grid points, with $\Delta x=\Delta y=0.04\,c/\omega_{pe}$ and 625 particles per species per cell, and a time-step $\Delta t=0.01/ \omega_{pe}$, a realistic mass ratio $m_p/m_e=1836$, and both the core and beam populations implemented as drifting Maxwellians (since the shape of the velocity distribution is not of interest in our analysis)\footnote{In the solar wind, suprathermal tails of Kappa distributions \cite{Pierrard-Lazar-2010} reduce the beam-core anisotropy, inhibit the electron-beam instability, and can even switch to the unstable regime of $L$ or $LB$ modes \cite{Lazar-etal-2023b}. 
This explains why radio emissions require higher beam speeds to compensate for the effects of suprathermal electrons \cite{Li-Cairns-2013, Li-Cairns-2014}.}.
We assume moderate magnetic field intensities, i.e., $\omega_{pe} /|\Omega_e| = 100$, as in \citeA{Kasaba-etal-2001, Rhee-etal-2009} and \citeA{Lazar-etal-2023a}, and also consistent with data linked to plasma sources of solar radio emissions \cite{Ergun_1998, Bale-etal-1999,Pulupa-etal-2010}.

%%%%%%%%%%%%%%%%%%%%
\section{Cases of interest} 
%%%%%%%%%%%%%%%%%%%5

We reconsider the three electron beam plasma configurations in \citeA{Lazar-etal-2023a}, all leading to primary waves with upshifted but mainly downshifted frequencies: two configurations with a single beam of electrons and one with counter-beams, whose parameters are summarized in Table~\ref{t1}.

%%%%%%%%%%%%%%%%%%%%%%
\subsection{Case~1} 
%%%%%%%%%%%%%%%%%%%%%%

\begin{figure*}[t!]
   \centering   
   \includegraphics[width=\textwidth,trim={0.3cm 3.5cm 0.3cm 3.5cm},clip]{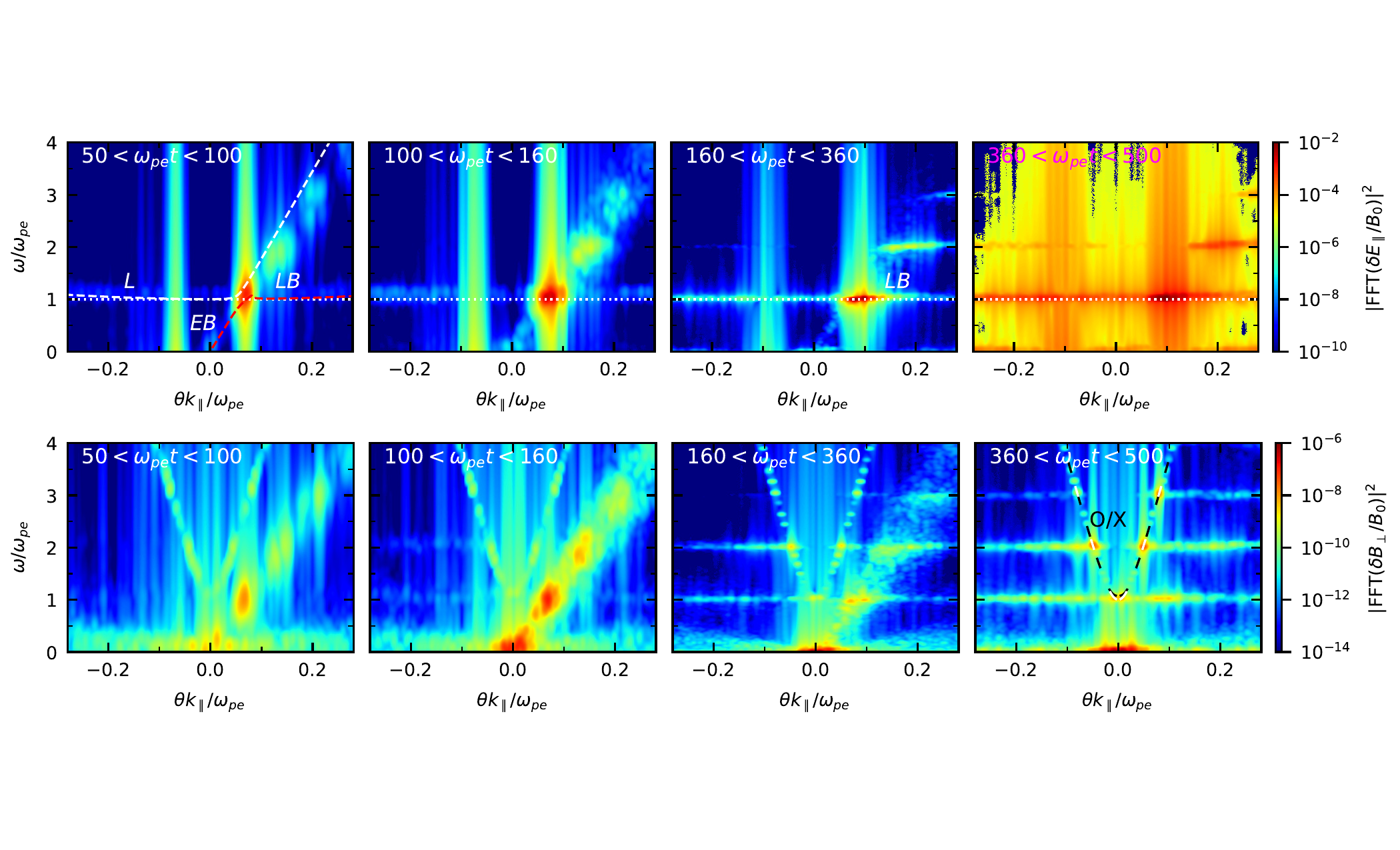}
\begin{internallinenumbers}
   \caption{\small Case 1: Wave energy density spectra (normalized) computed with spatio-temporal FFTs of the parallel electric field (upper) and perpendicular magnetic field (lower) components, as functions of frequency and parallel wavenumber, for different temporal intervals at $k_\perp=0$.}\label{f1}
\end{internallinenumbers}
   \end{figure*}

\begin{figure*}[t!]
   \centering   
   \includegraphics[width=\textwidth,trim={4.4cm 0cm 4.5cm 0cm},clip]{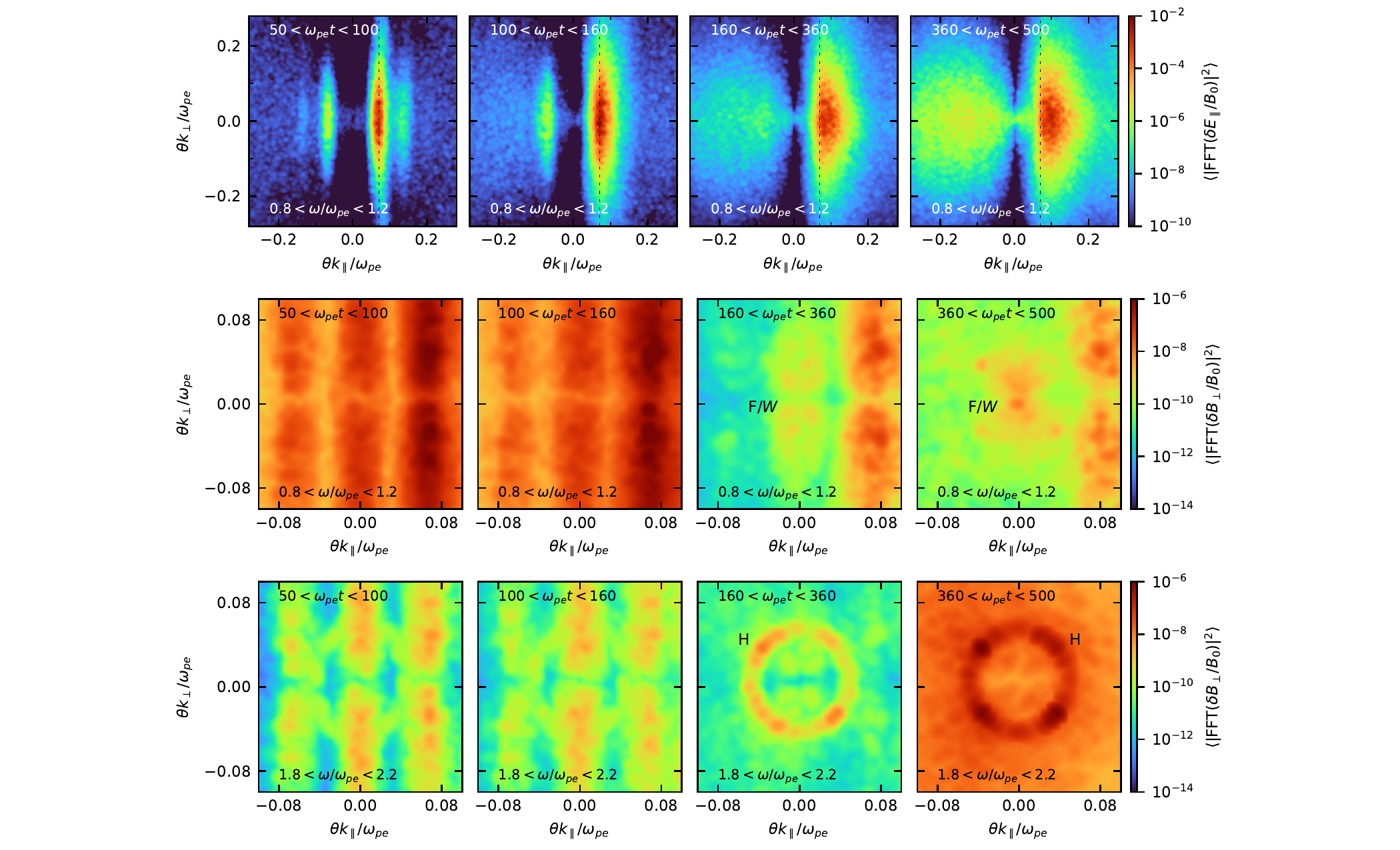}
\begin{internallinenumbers}
   \caption{\small The same wave energy density spectra (normalized) as in Figure~\ref{f1}, but as functions of parallel and perpendicular wavenumbers (normalized), and averaged in the frequency intervals $0.8< \omega/\omega_{pe} < 1.2$ (top and middle panels), and $1.8< \omega/\omega_{pe} < 2.2$ (bottom panels).}\label{f2}
\end{internallinenumbers}
   \end{figure*}

\begin{figure}[t!]
   \centering   %\includegraphics[width=0.48\textwidth]{kt_case1_all.pdf} \\
   \includegraphics[width=0.79\textwidth]{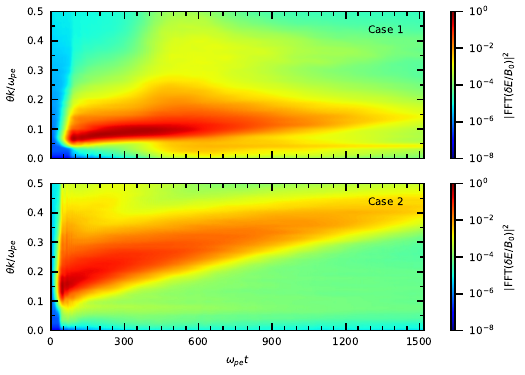}
\begin{internallinenumbers}
   \caption{\small Temporal evolution of $|{\rm FFT}(\delta E/B_0)|^2$ for case~1 (upper) and case~2 (lower), each of them color-coded and as a function of total wavenumber $k=(k_\parallel^2 + k_\perp^2)^{1/2}$.}\label{f3}
\end{internallinenumbers}
   \end{figure}

We first refer to PIC simulations of plasma emissions involving a single electron beam and start with case~1 in Table~\ref{t1}.
The upper panels in Figures~\ref{f1}--\ref{f3} display (normalized) wave energy density computed as spatio-temporal Fast Fourier Transforms (FFTs) of the parallel electric field component, $|{\rm FFT}(\delta E_\parallel/B_0)|^2$.
The wave dispersion curves derived (for initial conditions) from linear theory \cite{Cairns-1989, Lazar-etal-2023a} are also shown: the $L$ waves, with the forward ($k>0$) $L_+$ branch and the backward ($k<0$) $L_-$ branch, with white dashed lines; the almost linear branch of $EB$ waves at low $k$, continued after the shoulder (at $\sim \omega_{pe}$) with the $LB$ branch at higher $k$, with red dashed lines; and for reference, the white dotted line indicates $\omega=\omega_{pe}$. 
%Details of the corresponding dispersion relations, as well as the nature of the wave modes, can be found in \ref{linear-theory}.
For the initial condition in case~1, theory predicts an instability of $EB$ modes with maximum growth rate (Table~\ref{t1}) corresponding to a slightly downshifted (normalized) frequency $\omega_m / \omega_{pe} = 0.951 \lesssim 1$ and a (normalized) wavenumber $\theta_e k_m/ \omega_{pe} = 0.006$ \cite{Lazar-etal-2023a}.
In numerical simulations, as well as in real plasma systems, in addition to this fastest-growing mode, all neighboring modes with finite growth rates are subsequently excited.
The primary excitations actually have a spectrum of finite widths in frequency and wavenumber, around the fastest-growing $EB$ mode. 
The primary wave excitations are therefore found in a band of lower or higher frequencies, which means that the primary spectra contain both downshifted and upshifted frequencies.

The simulated spectra show the transition along the red-dashed dispersion curve in Figure~\ref{f1}, from the primary excitations, predominantly $EB$ modes, to $LB$ excitations\footnote{Also known as electron-acoustic modes, which are however rapidly damped in the absence of a high temperature contrast between the electron populations \cite{Sauer-etal-2019}.} \cite{Sauer-etal-2019, Lopez-etal-2020}. 
From left to right panels, we find a transition from $\omega / \omega_{pe} \lesssim 1$ and low wavenumbers $\theta_e k/ \omega_{pe} \lesssim 0.1$, to $LB$ waves at $\omega / \omega_{pe} \gtrsim 1$ and larger wavenumbers $\theta_e k/ \omega_{pe} \gtrsim 0.1$.
%The Langmuir beam branch is also known as the branch of electron-acoustic modes \cite{Sauer-etal-2019}, which helps to differentiate, especially when it remains predominantly below the plasma frequency (we will see in case 2 below).
Correspondingly, the wave spectra in Figure~\ref{f2}, upper panels, show maximum intensities evolving toward larger (parallel) wavenumbers, compared to the dashed line marking the initial maximum.
%The asymmetry of these spectra with respect to the positive (forward) and negative (backward) directions of $k_\parallel$ is highlighted by extracting the relevant frequencies in the vicinity of $\omega_{pe}$, namely, in the interval $0.8< \omega/\omega_{pe} < 1.2$ for $L$ waves and $0.8< \omega/\omega_{pe} < 2.2$ for $T$ waves; see also \citeA{Lee-etal-2019}.
These spectra are plotted as a function of parallel and perpendicular wavenumbers by averaging the spectra in Figure~\ref{f1} around the plasma frequency, specifically over the interval $0.8< \omega/\omega_{pe} < 1.2$.
For $\omega_{pe}t > 300$, these high-frequency spectra are dominated by $LB$ excitations, as the culminating stage of the transition from downshifted to upshifted excitations.
Higher harmonics of $LB$ waves are also excited at $\sim 2\omega_{pe}$ and $\sim 3\omega_{pe}$ \cite{Gaelzer-etal-2002, Yoon-etal-2003, Sauer-Sydora-2012}, with higher intensities in the forward direction. %, and can contribute to the generation of higher harmonics of radio $T$ waves.
Upshifted $LB$ waves conform to the standard plasma model of radio emission and can engage in wave-wave interactions (with, e.g., $S$ and $L$ waves) to ultimately produce $T$ radio waves \cite{Melrose_2008}, as explained in Section~\ref{sec-1}. %, both the F and H emissions obtained in Figure~\ref{f1}, third and fourth panels lower.

%   \begin{figure}[t!]   \centering   \includegraphics[width=0.48\textwidth]{kxt_case1.pdf}
%\begin{internallinenumbers}
%   \caption{\small Time evolution of $|{\rm FFT}(\delta E_\parallel/B_0)|^2$ as a function of $k_\parallel$ (upper), and $|{\rm FFT}(\delta B_\perp/B_0)|^2$ as a function of $k_\perp$ (lower), each of them color codded.}\label{f4}
%\end{internallinenumbers}
%   \end{figure}

The spectra of (normalized) magnetic wave energy density, computed as $|{\rm FFT}(\delta B_\perp/B_0)|^2$, are displayed in Figure~\ref{f1}, lower panel, and Figure~\ref{f2}, middle and bottom panels.
In the middle panels, we tracked the F~$\to T$ component, averaging the spectra in the interval $0.8< \omega/\omega_{pe} < 1.2$, and in the bottom panels the H~$\to 2T$ component in the interval $1.8< \omega/\omega_{pe} < 2.2$.
%The 2D wave spectra in Figure~\ref{f2}, upper panels, confirm the 
More intense $T$ wave spectra are also obtained here after $\omega_{pe} t = 300$, both the fundamental (F) and higher (H) harmonics with peak intensities for oblique propagation, and in agreement with the arguments from \citeA{Sauer-Sydora-2012} and \citeA{Sauer-etal-2019}.
%Their similar simulations evidenced subsequent relaxation (by heating and deceleration) of the beam to a plateau, and associated $LB$ excitations as pump waves decaying into low-$k$ $L$ waves and $S$ waves (three-wave process).
Moreover, the evolutions of the electron velocity distributions shown by \citeA{Thurgood-Tsiklauri-2015} and \citeA{Sauer-etal-2019} for similar (initial) setups suggest that primary excitations have an indirect, rather than a direct, involvement in the radiative process. 
Specifically, growing fluctuations are effective in relaxing and flattening the beam, as proved both in the absence \cite{Thurgood-Tsiklauri-2015, Henri-etal-2019, Lee-etal-2019} and in the presence of a magnetic field  \cite{Kasaba-etal-2001, Sauer-etal-2019, Lee-etal-2022}.
Such that the beam relaxes to a plateau-on-tail distribution of lower density but hotter, which can trigger $L$ or $LB$ waves with higher frequencies, predominantly above the plasma frequency, $\omega \geqslant \omega_{pe}$.
$LB$ waves with large enough $k$ can decay into forward propagating $S$ waves and $L$ waves with low-$k$ (optical range), $LB \to L+S$, a wave-wave decay also known as Langmuir condensation \cite{Lee-etal-2019}.

In addition, \citeA{Sauer-etal-2019} invoked a mechanism distinct from those in the standard plasma model, namely, linear mode conversion (LMC) of low-$k$ $L$ waves to $T$ waves.
Originally predicted for non-magnetized plasmas, when coupling of longitudinal $L$ waves and radio $T$ waves is facilitated by the presence of inhomogeneities \cite{Field-1956}, LMC is also invoked for type II and type III bursts \cite{Lin-etal-1981, Melrose1985, Cairns2011}, and in planetary magnetospheres \cite{Yoon1998, Menietti2009, Schleyer-2014}.
LMC operates when dispersion curves, wave frequency vs. wavenumber ($\omega$ vs. $k$), approach or even cross each other \cite{Sauer-etal-2019}.
In overdense plasmas ($\omega_{pe} > |\Omega_e|$), the dispersion curves of the $L$ mode and the free EM modes, left-handed (LH) and right-handed (RH) modes, cross each other for parallel propagation, but their fields remain uncoupled. 
Instead, oblique modes can couple, that is, Langmuir/Z ($LZ$) mode with, respectively, the ordinary ($O$) and extraordinary ($X$) modes (corresponding to the LH and RH modes in the parallel direction), and their dispersion curves indeed split, proving conversion from one mode to another around the crossing point; see Figure~9 in \citeA{Sauer-etal-2019}.
The initial plasma is homogeneous, but if density gradients are generated later in the simulation, the parallel $L$ modes may also undergo LMC into radio emissions \cite{Field-1956, Volokitin_2020}.
For parallel propagation in Figure~\ref{f1} (last lower panel), the $O$ and $X$ modes are also indicated with dashed lines, white and black, respectively. 
In overdense plasmas these two modes overlap, and only near $\omega_{pe}$ the $O$ mode is slightly downshifted \cite{Lee-etal-2022}, which can explain the observed F emission (as an $O$ mode) with a weak partial polarization and an almost depolarized second H component \cite{Sasikumar_2013, Pulupa_2020}.  
%The presence of enhanced ion-density fluctuations at the expected $k_S \gtrsim k_{LB}$ is also confirmed.

In our Figures~\ref{f1} and \ref{f2}, the spectra in the fourth upper panels are consistent with both scenarios, showing excitations of $LB$ and $L$ waves at $\sim \omega_{pe}$, but also $S$ waves at much lower frequencies, along a wide range of wavenumbers, including both forward and backward propagation. 
Also, the 2D spectra in the last two upper panels of Figure~\ref{f2} confirm the extension of the $L$ modes, including the oblique ones, towards the lower $k_\parallel$. 
It is very likely that such bursty spectra of excitations $LB$, $L$ and $S$ lead to multiple mechanisms capable of producing radio emissions, as discussed below.
The upper panel of Figure~\ref{f3} shows the full evolution of the wave energy density, that is, $|{\rm FFT}(\delta E/ B_0)|^2$ as a function of the total wavenumber, from our long-run simulation ($\omega_{pe} t_{\rm max} = 1500$). 
The transition from low-$k$  excitations ($EB$ modes) to those with larger $k$ ($LB$ waves) starts early, before $\omega_{pe}t = 200$, and after $\omega_{pe}t = 350$ intense primary $LB$ waves occur at even larger $k$. 
After $\omega_{pe}t \simeq 400$, clear signatures of the decay of $LB$ waves into daughter waves of higher and lower wavenumbers ($S$ waves and $L$ waves, respectively) are observed, with the branch of the latter distinctly separating at very low $k \to 0$.
%
%Compared to the simulations in \citeA{Sauer-etal-2019}, the nonlinear decay in our case~1 seems to occur earlier, possibly due to more intense fluctuations generated by a (slightly) denser and more energetic beam.

The spectra of $T$-modes in Figures~\ref{f1} and \ref{f2}, lower panels, are complex. 
Free radio waves can escape and propagate in the ambient plasma with sufficiently isotropic spectra, and can thus be identified within circular shapes \cite{Lee-etal-2019, Lee-etal-2022}, such as the intense second H emission in Figure~\ref{f2}, lower panels.
%, for both forward ($k>0$) and backward ($k<0$) propagations. 
A typical quadrupole pattern (oblique peaks) of F emission at plasma frequency ($\sim \omega_{pe}$) and low $k$ generally connects with the peaks of H emissions at larger $k$.
The circular shapes of their maxima are more specific to weakly or nonmagnetized plasmas \cite{Lee-etal-2022}.
Again, our spectra in Figure~\ref{f1}, lower panels, suggest that harmonics of the $LB/L$-modes undergo LMC into the corresponding radio H waves, where their dispersion curves cross each other, at, e.g., $2 \omega_{pe}$, $3 \omega_{pe}$, etc.
%The most intense are the $2T$-free modes, compared to F and $3T$-waves, in the spectra from Figures~\ref{f1} and \ref{f2}, last lower panels.
In addition, the 2D spectra in Figure~\ref{f2}, the last two lower panels, confirm the oblique propagation maxima for the radio components F and H.
However, given the bursty spectra of primary and parametric excitations in, e.g., Figure~\ref{f1}, last upper panel, not only one but several distinct mechanisms are expected to be at the origin of radio emissions: wave-wave decays, e.g., $LB \to  S + T$, and coalescence, e.g., $L_- + S_+  \to T$, $LB + L_- \to 2T$, etc. (where subscripts $+$ and $-$ indicate forward and backward propagation, respectively), but also the LMC discussed above. %\footnote{Despite some conclusions of previous lower resolution simulations \cite{Kasaba-etal-2001}.}. 
The high intensity of the excitations corresponding to these modes (but also of the quasi-thermal noise in between) proves that the resonance conditions are fully satisfied, not only between the electron beam and the primary excitations, but also in the wave-wave interactions generating the secondary excitations.
In the magnetic wave energy spectra (out-of-plane component, $B_\perp$), lower panels of Figures~\ref{f1} and \ref{f2}, we can also recognize the EM electron beam modes excited in the early stage by the Weibel or filamentation-like instability \cite{Lazar-etal-2010}, with highly oblique or nearly perpendicular propagation \cite{Karlicky-2009, Thurgood-Tsiklauri-2015}. 
The radio waves apparently emerge later in these spectra, for, e.g., $\omega_{pe} t > 150$, and with maxima for less oblique wave vectors with comparable components, $k_\parallel \sim k_\perp$. 
The radio component F can be difficult to distinguish due to the Weibel ($W$) modes, the overlap of the two excitations being indicated by F/$W$ (Figure~\ref{f2}).

%%%%%%%%%%%%%%%%%%%%%%%%
\subsection{Case~2}
%%%%%%%%%%%%%%%%%%%%%%%%%%

\begin{figure*}[t!]
   \centering   
   \includegraphics[width=\textwidth,trim={0.3cm 3.5cm 0.3cm 3.5cm},clip]{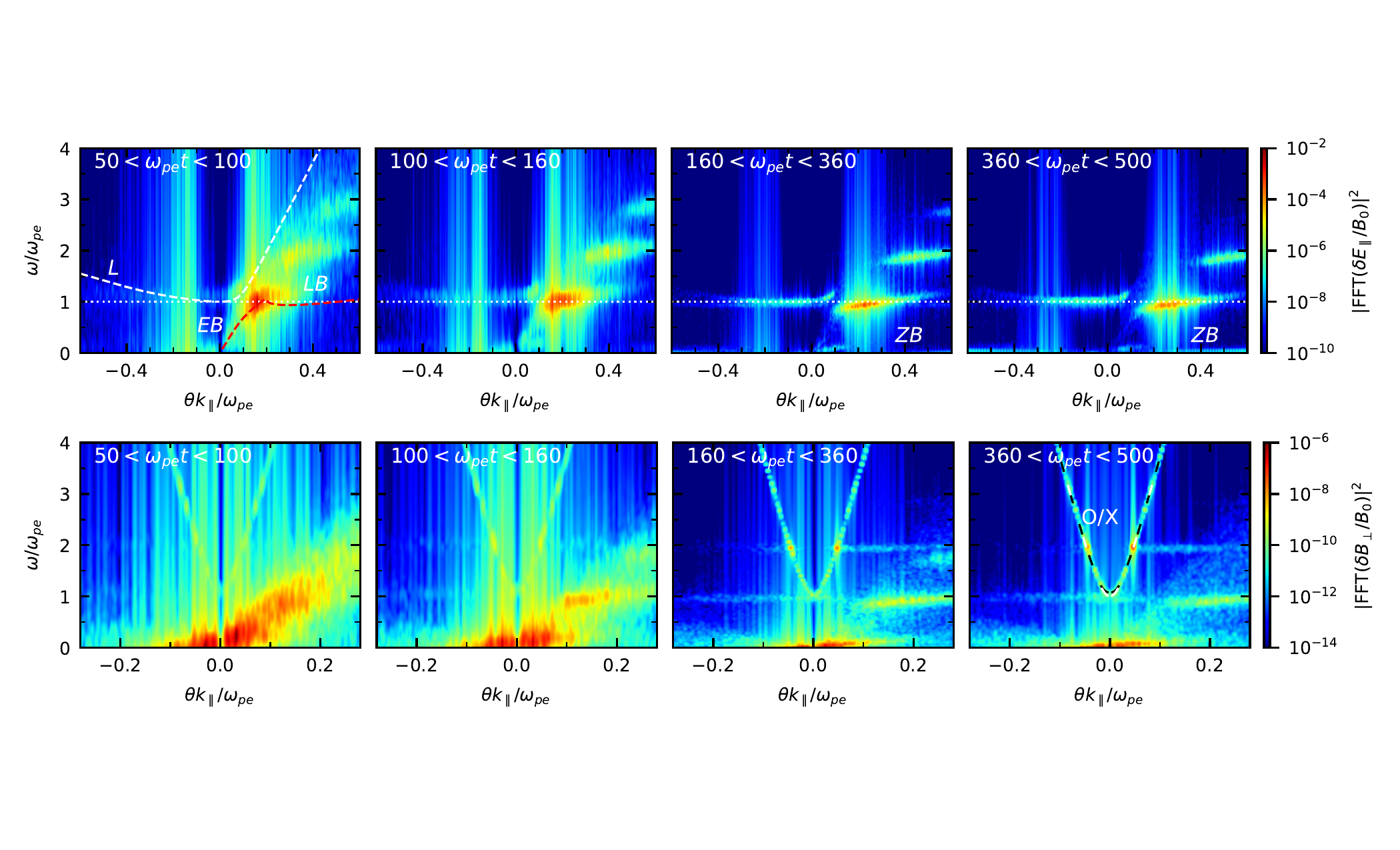}
\begin{internallinenumbers}
    \caption{\small Case 2: Wave energy density spectra (normalized) computed with spatio-temporal FFTs of the parallel electric field (upper) and perpendicular magnetic field (lower) components, as functions of frequency and parallel wavenumber, for different temporal intervals at $k_\perp=0$.}\label{f4}
\end{internallinenumbers}
   \end{figure*}

\begin{figure*}[t!]
   \centering   
   \includegraphics[width=\textwidth,trim={4.4cm 0cm 4.5cm 0cm},clip]{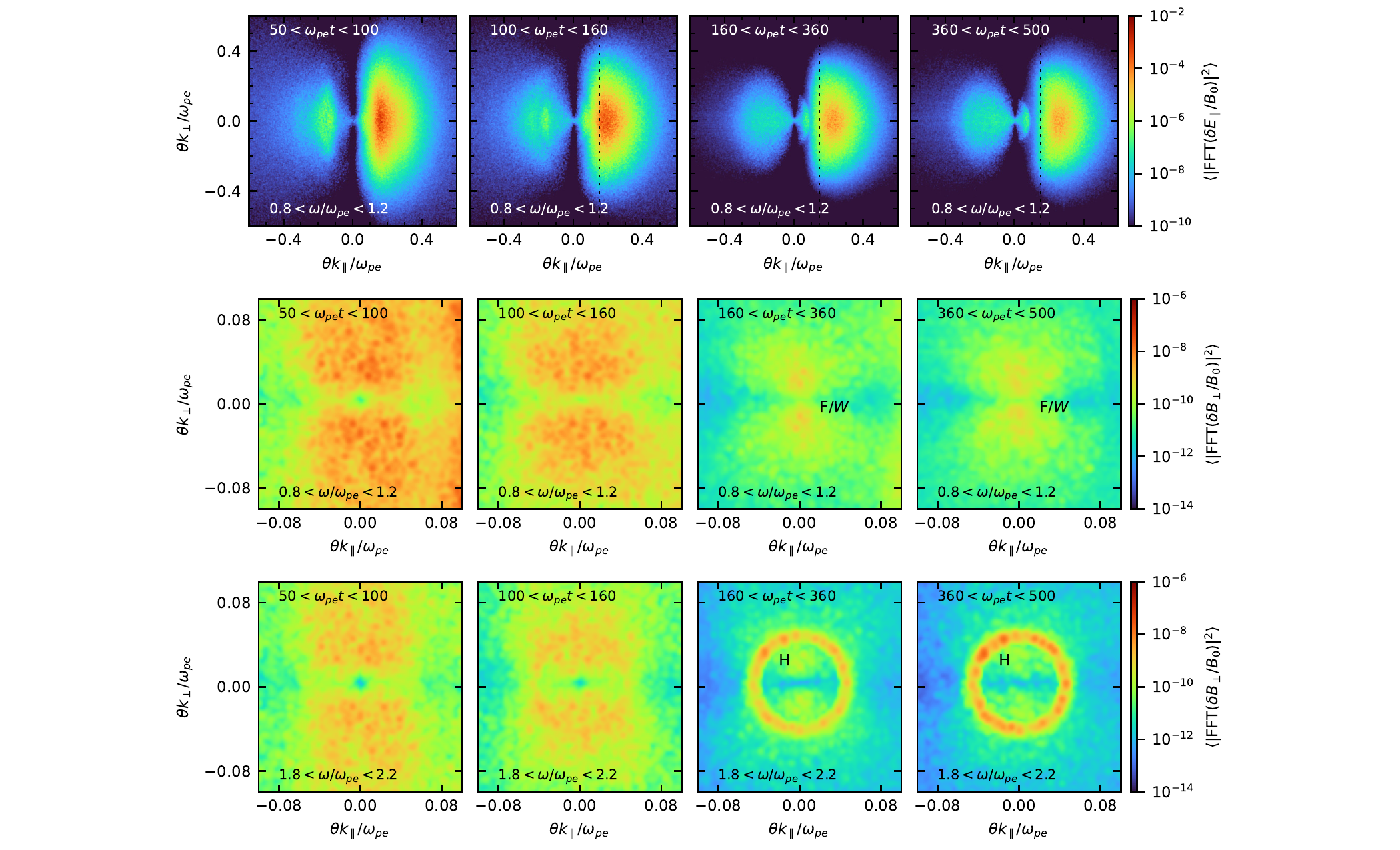}
\begin{internallinenumbers}
   \caption{\small The same wave energy density spectra (normalized) as in Figure~\ref{f4}, but as functions of parallel and perpendicular wavenumbers (normalized), and averaged in the frequency intervals $0.8< \omega/\omega_{pe} < 1.2$ (top and middle panels), and $1.8< \omega/\omega_{pe} < 2.2$ (bottom panels).}\label{f5}
\end{internallinenumbers}
   \end{figure*}

In cases~2 and 3 we change the (initial) setups of the electron beam plasma, see Table 1, and downshifted primary excitations are involved again.
In case~2, we deal with a single beam as in case~1, but denser ($n_b/n_e = 0.05$) and with a lower relative beam speed ($U_b/\theta_e =8$).
The corresponding FFT spectra obtained in case~2 for the wave energy densities are displayed in Figures~\ref{f4} and \ref{f5}.
These results correspond to the same time interval that was found relevant for the generation of radio waves in case~1. 
Returning to Figure~\ref{f3}, the lower panel shows the temporal evolution of $|{\rm FFT}(\delta E/B_0)|^2$ in case 2, throughout the entire interval of our extended run ($\omega_{pe}t_{\rm max} = 1500$).

\begin{figure*}[t!]
   \centering   
   \includegraphics[width=\textwidth]{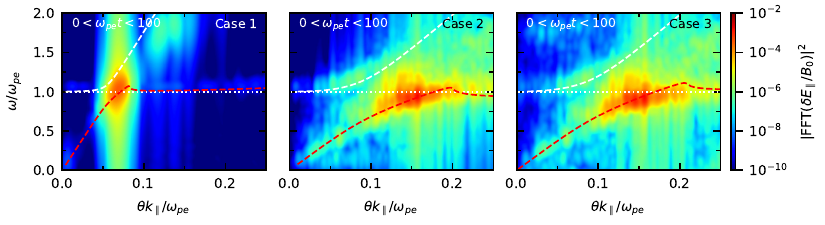}
\begin{internallinenumbers}
   \caption{\small Comparison of the early primary spectra with details of the downshifted excitations.}
   \label{f5-6}
\end{internallinenumbers}
   \end{figure*}

Looking for explanations for the differences between the spectra in cases~1 and 2, we first turn to the predictions of linear theory for the primary excitations; see in \citeA{Lazar-etal-2023a} their Figures~1 and 2, and the related comments.
The fastest growing (with maximum growth rate) is also an $EB$ mode with $\omega_{\rm m}/ \omega_{pe} = 0.897$, slightly lower than in case~1 (Table~\ref{t1}), while the corresponding wavenumber becomes much higher $k_{\rm m} \theta_e/ \omega_{pe} = 0.142$ \cite{Lazar-etal-2023a}.
These values prove that the instabilities of the $EB$ modes ($\omega \simeq kU_b$) are still kinetic or Landau-resonant (as long as their frequency remains close to $\omega_{pe}$).  
A slightly lower frequency also implies that the early spectrum of primary excitations appears to be more attached to the $EB$ branch and therefore more downshifted along this branch, compared to case~1.
This is confirmed by the early spectra of primary excitations compared in Figure~\ref{f5-6}. 
In case 2, the primary excitations are more extended along the EB branch, being visibly dominated by the downshifted frequencies below the plasma frequency ($\omega_{pe}$), indicated by the white dotted line.

\begin{figure*}[t!]
   \centering   
   \includegraphics[width=\textwidth,trim={0.3cm 2.5cm 0.3cm 3cm},clip]{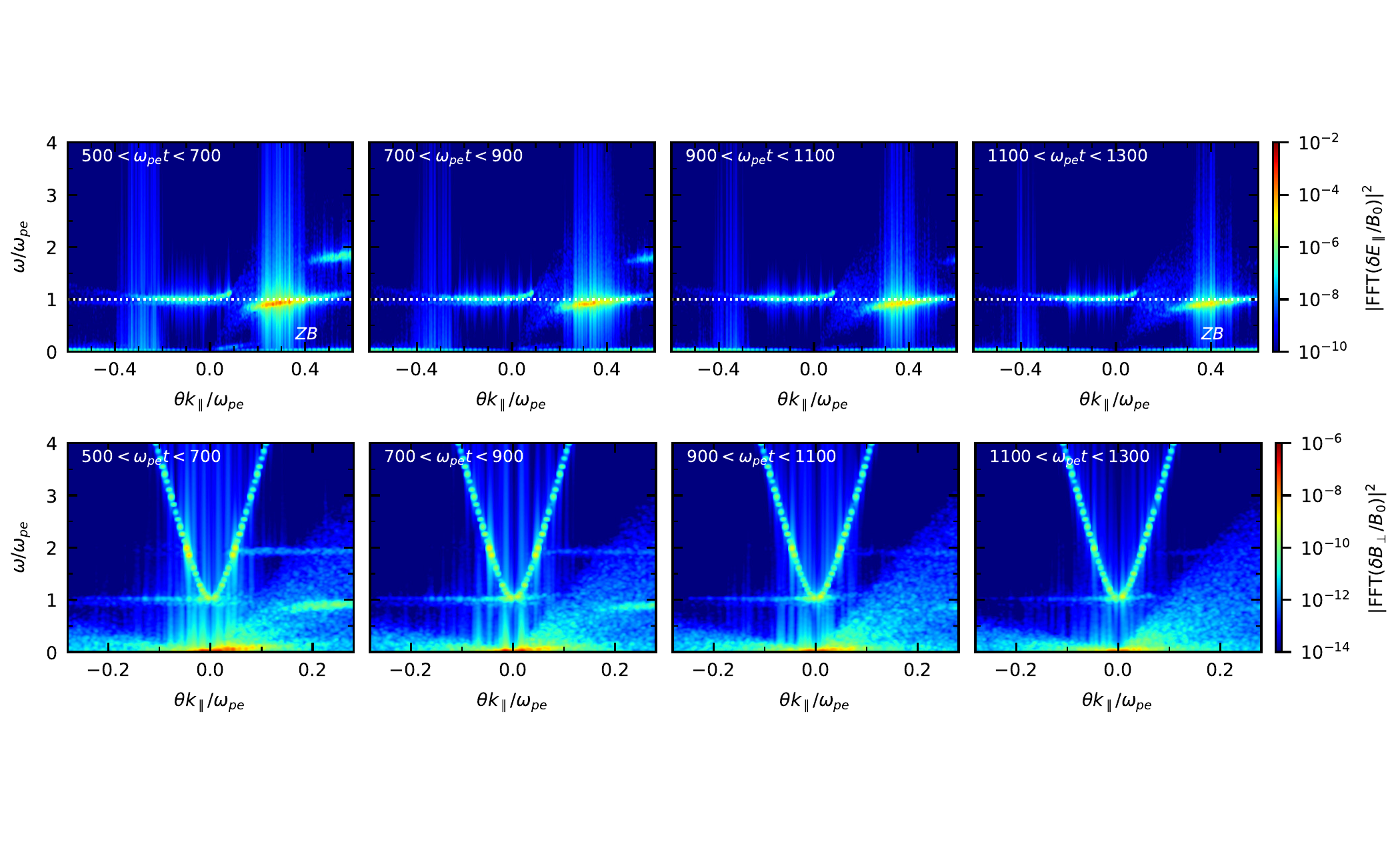}
\begin{internallinenumbers}
   \caption{\small Case 2: The same as in Figure~\ref{f4}, but for extended temporal intervals (indicated in each panel).}
   \label{f6}
\end{internallinenumbers}
   \end{figure*}

Initial differences also have consequences later in time when these excitations evolve, also toward larger wavenumbers (upper panels in Figure~\ref{f5}), and contrast to those in case~1; see also the contrast between lower and upper panels in Figure~\ref{f3}. 
The beam undergoes a similar relaxation to a plateau-on-tail \cite{Thurgood-Tsiklauri-2015}, which again determines the spread of the primary wave maxima along the $LB$ branch (red dashed line) with frequencies predominantly above $\omega_{pe}$, see upper panels of Figures~\ref{f4} and \ref{f5}.
However, these $LB$ excitations have higher wavenumbers, as also proven by the lower panel in Figure~\ref{f3}, which seems to markedly affect the resonance conditions for the further wave-wave decays. 
The $LB$ waves have high enough frequencies $\omega \gtrsim \omega_{pe}$, but the lower panel of Figure~\ref{f3} shows no signature of an effective resonant decay as in case~1. 
Also, the upper panels of Figure~\ref{f4}, but also those of the subsequent intervals in Figure~\ref{f6}, do not present explosive spectra specific to resonant wave interactions (as in Figure 1, last upper panel).
Also, the lower panel of Figure~\ref{f3} does not show the same prominent decay as in the upper panel in case~1. 
However, in Figure~\ref{f4} one can still distinguish parametric (secondary) excitations of much lower intensities, forward propagating $S$ waves with large $k$ and backscattered $L_{-}$ waves with small $k$.
%Higher intensities of $L$ and low-frequency $S$ waves can be noticed in the upper (last) panels of Figure~\ref{f4}. %, but also all the later time sequences in Figure~\ref{f8a}, upper panels, in Appendix~\ref{ap-b}.
In turn, the primary and parametric excitations can couple, leading to radio emissions, both the F component, via $L_- + S_+ \to T$, and the second harmonic H, via $LB + L_{-} \to 2T$.
On the other hand, primary $LB$ waves can also decay into a $S$ wave with $k_S \simeq k_L$, and a forward propagating $L_+$ wave with low $k_L$ in the optical range ($k_L c \sim \omega_{pe}$) via $LB \to S_+ + L_{+}$.
These secondary excitations can be identified in Figure~\ref{f4}, easier to distinguish than in case 1 due to the less bursty and noisy spectra.
As discussed in case~1, $L$ waves can be responsible for the generation of $T$ waves of radio F emission through LMC, and if primary $LB$ waves are sufficiently intense, it becomes also possible to decay directly as $LB \to S + T$.

\begin{figure*}[t!]
   \centering   
   \includegraphics[width=\textwidth,trim={0.2cm 1.7cm 0.2cm 1.5cm},clip]{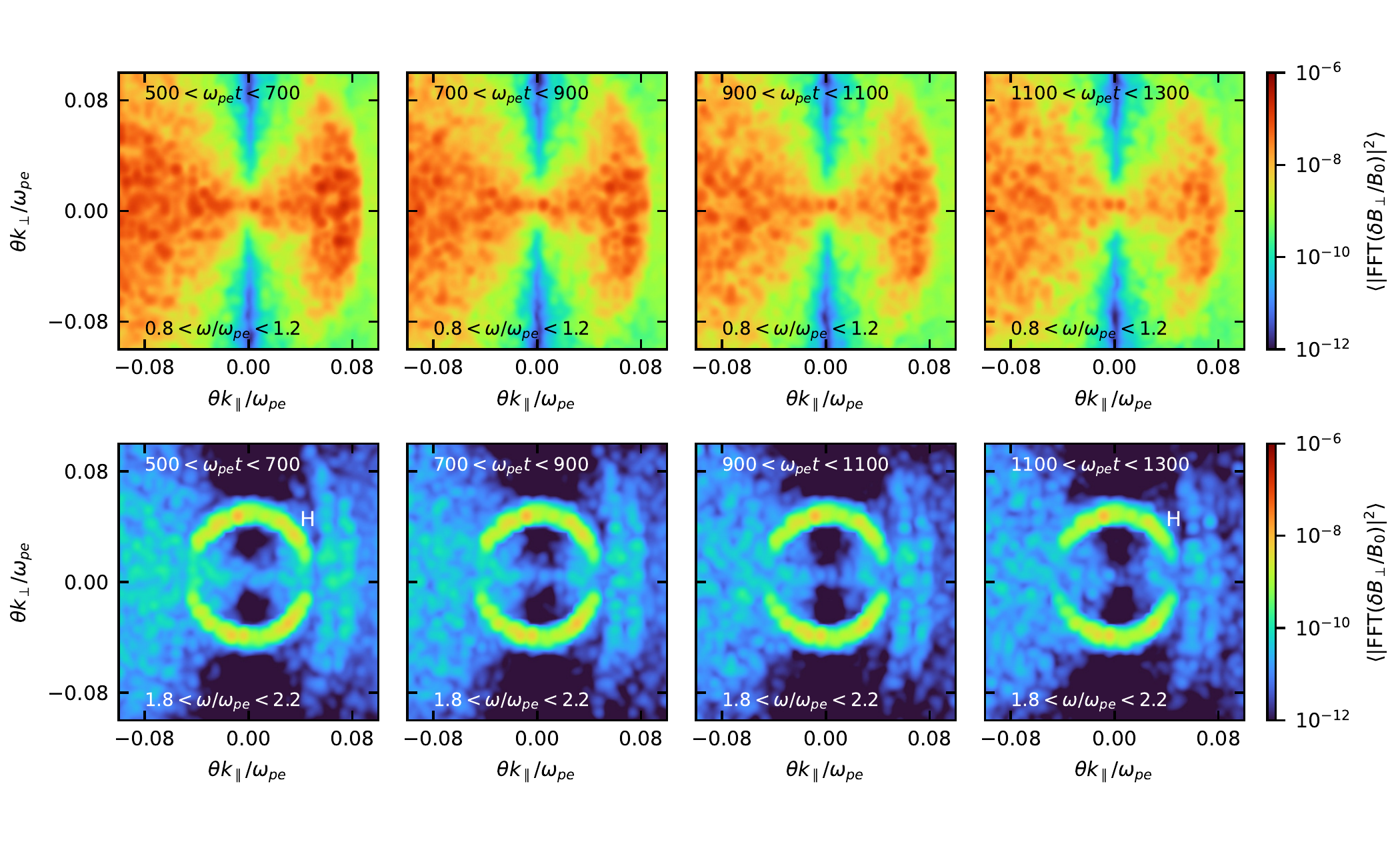}
\begin{internallinenumbers}
   \caption{\small Case 2: The same as middle and bottom panels of Figure~\ref{f5}, but for extended temporal intervals (indicated in each panel).}
   \label{f7}
\end{internallinenumbers}
   \end{figure*}

The decay processes of the primary excitations involve a weak wave-wave resonance this time, which makes the intensities reached by the secondary (or parametric) excitations, for example $S$ and $L$ waves, much lower than in case~1.
This also affects the processes of coalescence (or coupling) of primary and secondary excitations, and can explain the very low levels of F emission (produced by LMC or a coalescence like $L +S \to T$), barely distinguishable in Figure~\ref{f4}, lower panels, and Figure~\ref{f5}, middle panels (possibly mixed with $W$ excitations).
Instead, the second H radio waves remain more prominent, e.g. in the bottom last panel in Figure~\ref{f5}, even when the primary $LB$ excitations tend to drop below the plasma frequency, in the upper last panel of Figure~\ref{f4}. 
These modes have frequencies $\omega <\omega_{pe}$ and we call them beam-induced excitations of the $Z$ mode ($ZB$) \cite{Lee-etal-2022}, to distinguish them from $LB$ excitations with $\omega>\omega_{pe}$.

In the longer runs in Figure~\ref{f6}, the primary $ZB$ waves are further attenuated in intensity, and their wide wavenumber band drops below the plasma frequency. 
Rather, these appear to be electron-acoustic modes excited by a relaxed and thermalized electron beam, as a plateau on the tail, hotter than the core \cite{Sauer-etal-2019}.
The primary excitations become predominantly downshifted $ZB$ waves (see the upper last panel of Figure~\ref{f4}), while the second H radio emission remains prominent (compared to previous temporal intervals).
In addition, the radio emission shows an isotropization tendency, unlike the more quadrupolar spectrum in case~1.
These features strongly suggest that H emission is the result of the coalescence of primary waves, upshifted $LB$ waves but also downshifted $ZB$ excitations, with parametric $L_{-}$  modes, namely, via $LB + L_{-} \to 2T$ and $ZB + L_{-} \to 2T$, respectively.
More direct confirmations of the involvement of downshifted $ZB$ primary waves are provided by the spectra of subsequent time intervals in Figures~\ref{f6} and \ref{f7}. 
The upper panels of Figure~\ref{f6} show fully downshifted $ZB$-type primary excitations, and correspondingly, a less intense (compared to earlier times) but still appreciable second H radio component is obtained in Figure~\ref{f7}.
This is the first concrete evidence for direct involvement of downshifted primary excitations in the generation of radio emissions.
In this case, it occurs late (on large time scales) when the beam is already flattened.
In case 3 discussed in the following, we will see a faster and more pronounced involvement of the downshifted primary excitations.
We cannot rule out that similar processes also take place in case~1 as well, especially during the bursty spectra in the last upper panel of Figure~\ref{f1}, with primary excitations above and below the plasma frequency.
In the radio spectra in Figure~\ref{f2} the emission of the H component is also much more intense than the F component.

%%%%%%%%%%%%%%%%%%%%%%%%
\subsection{Case~3}
%%%%%%%%%%%%%%%%%%%%%%%%%%

In case~3, the plasma system is significantly different, considering two symmetric counterbeams of electrons which lead to major differences in the EM radio wave spectra, as well as in the underlying mechanisms.
Double or bidirectional beams, with the same properties as the beam in case~2, see Table~\ref{t1}, produce symmetric spectra of counter-propagating primary excitations (subscripts "+" for forward and "-" for backward propagation), predominantly downshifted with respect to plasma frequency; see also the last panel of Figure~\ref{f5-6}.
The fastest growing modes, both forward and backward propagating, are predicted by linear theory for initial conditions and remain similar to that of case~2, with a slightly lower frequency $\omega_ {\rm m}/ \omega_{pe} = 0.877$ on the branch $EB$ (red-dashed lines) and a wavenumber $k_{\rm m} \theta_e/ \omega_{pe} = 0.139$ \cite{Lazar-etal-2023a}.

%We use the same filtering method (explained in section 2.1), to extract the relevant frequencies of radio waves, whose contrasting spectra are shown in Figure~\ref{f5}, lower panels.

\begin{figure*}[t!]
   \centering   
   \includegraphics[width=\textwidth,trim={0.3cm 3.4cm 0.3cm 3.4cm},clip]{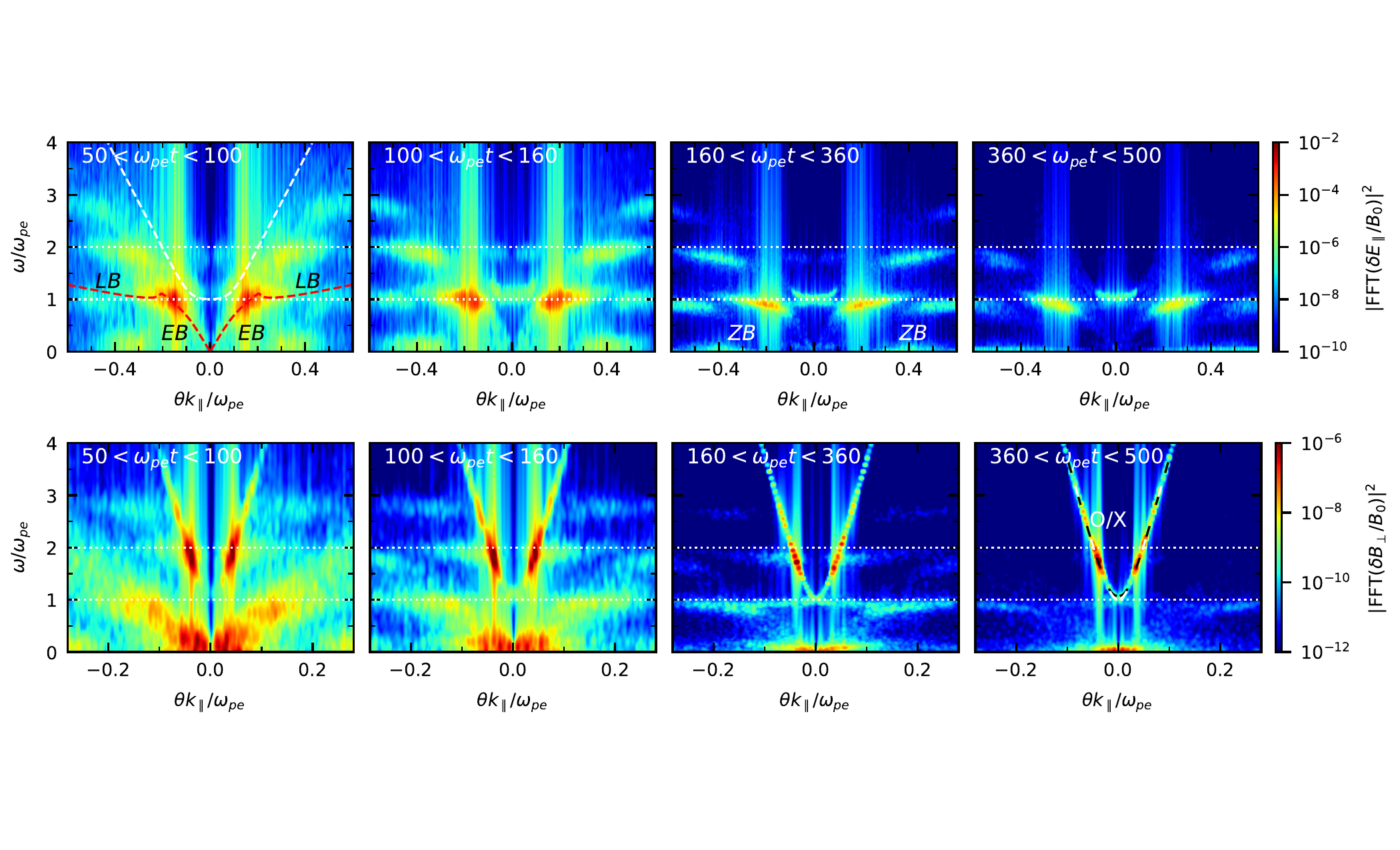}
\begin{internallinenumbers}
    \caption{\small Case~3: Wave energy density spectra (normalized) computed with spatio-temporal FFTs of the parallel electric field (upper) and perpendicular magnetic field (lower) components, as functions of frequency and parallel wavenumber, for different temporal intervals at $k_\perp=0$.}\label{f8}
\end{internallinenumbers}
   \end{figure*}

\begin{figure*}[t!]
   \centering   
   \includegraphics[width=\textwidth,trim={4.4cm 0cm 4.4cm 0cm},clip]{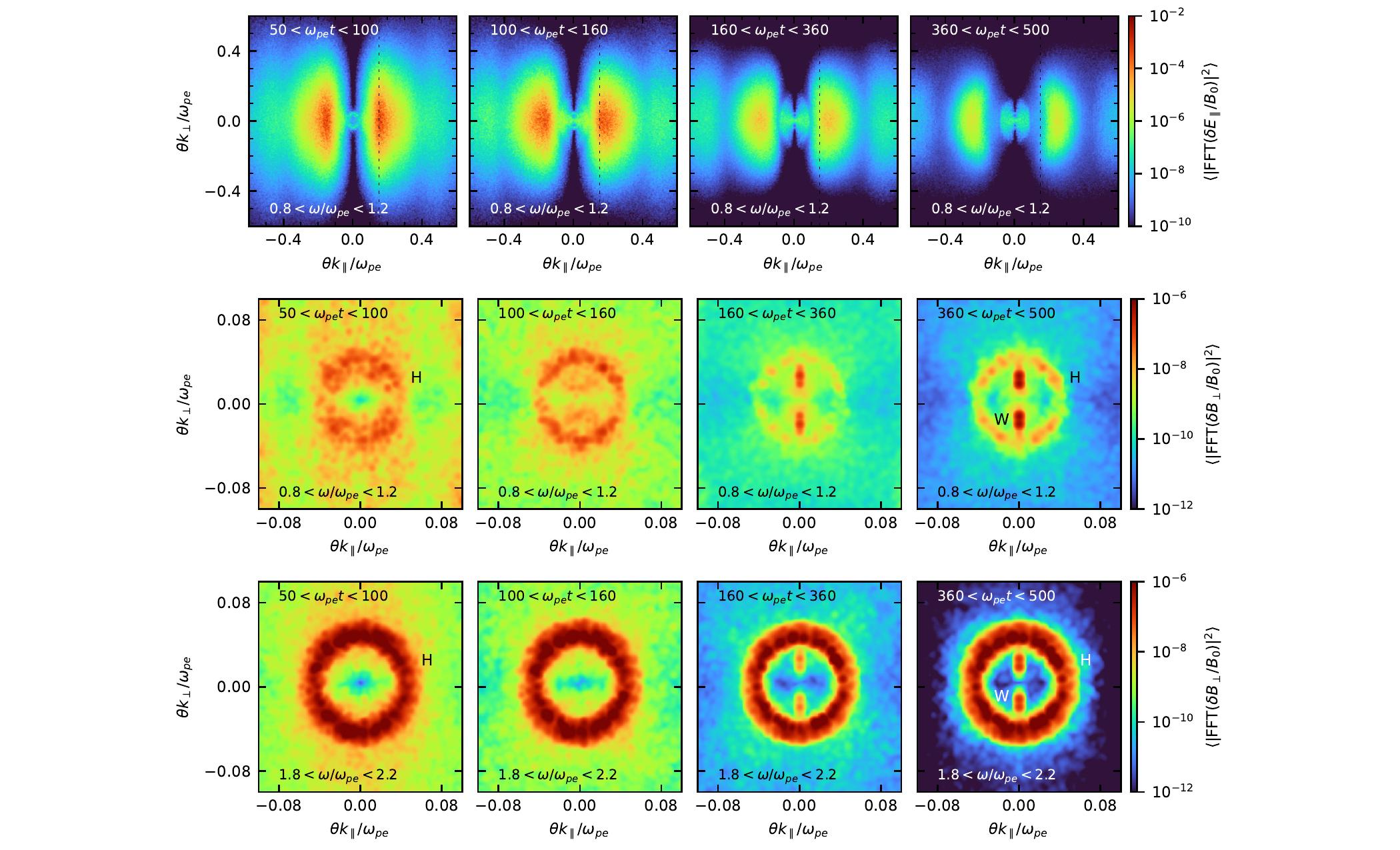}
\begin{internallinenumbers}
   \caption{\small The same wave energy density spectra (normalized) as in Figure~\ref{f8}, but as functions of parallel and perpendicular wavenumbers (normalized), and averaged in the frequency intervals $0.8< \omega/\omega_{pe} < 1.2$ (top and middle panels), and $1.8< \omega/\omega_{pe} < 2.2$ (bottom panels).}\label{f9}
\end{internallinenumbers}
   \end{figure*}

The simulated spectra of the primary excitations are shown in the upper panels of Figures~\ref{f8} and \ref{f9}, for the same time intervals as in Figure~\ref{f4}.
In the early phases, that is, $\omega_{pe}t < 160$ , both $EB$ modes with downshifted frequencies ($\omega \lesssim \omega_{pe}$) and $LB$ waves with upshifted frequencies ($\omega \lesssim \omega_{pe}$) are excited.
The lower panels in Figure~\ref{f8} and the middle and bottom panels in Figure~\ref{f9} prove very fast radio generation, most likely triggered by direct coupling (or coalescence) of primary excitations, i.e., $L_{+} + L_{-} \to 2T$, where here the symbol $L$ is used generically, for two counterpropagating modes of either $EB$ or $LB$ nature.
The resulting second harmonic (H$\to 2T$) is highly isotropic and very intense, in a broad frequency bandwidth centered on a predominantly downshifted frequency, i.e., $\omega \lesssim 2\omega_{pe}$.
These properties strongly suggest that at the origin of radio emission the coalescence of primary excitations with downshifted frequencies also contributes.
The frequency bandwidth of the second H radio emission is larger, about twice the bandwidth of the primary wave frequencies, as evidence that not only upshifted $LB$ waves participate in the coalescence leading to the second H radio emission, but also the downshifted $EB$ excitations.
Thus, the maximum values of the radio frequencies are approximately given by the maximum (upshifted) frequencies of the primary excitations, $\omega_{2T,max} \simeq 2 \omega_{LB,max} > 2 \omega_{pe}$, while the minimum radio frequencies are given by the downshifted primary excitations, $\omega_{2T,min} \simeq 2 \omega_{EB,min} < 2 \omega_{pe}$. 

%\begin{figure*}[t!]
%   \centering   \includegraphics[width=0.95\textwidth]{kxky_case3.pdf}\\
%   \vspace{-0.47cm}
%   \includegraphics[width=0.95\textwidth]{kxky_bz_case3.pdf}
%\begin{internallinenumbers}
%   \caption{\small Case 3: Time sequence of the integrated spectra of the parallel electric field component (upper) and the perpendicular magnetic field component (lower) as a function of parallel and perpendicular wavenumbers.}
%   \label{f9}
%\end{internallinenumbers}
%   \end{figure*}

In the later phases, i.e. $\omega_{pe}t > 160$, both primary excitations (forward and backward) decrease significantly in intensity but remain easy to identify because the spectra are less noisy. 
In the last two upper panels of Figures~\ref{f8} and \ref{f9}, their maxima have downshifted frequencies and move to higher wave numbers, typical for what we called $ZB$ mode excitations.
In Figures~\ref{f8} and \ref{f9}, lower panels, the H component of radio waves is produced with significant intensities and sufficiently isotropic spectra, mainly by the coalescence of the downshifted excitations, $ZB_+ + ZB_- \to 2T$. 
Their spectra are better outlined along the dispersion curves of free EM modes, and the maxima in the frequency band drop below $2 \omega_{pe}$.
This is another important proof of the direct involvement of primary downshifted excitations in the generation of EM radio waves.
It should be noted that excitations predominantly downshifted are signaled in this case much earlier ($\omega_{pe}t > 160$) than in case~2 ($\omega_{pe}t > 500$).
The excitations of two counter-beams of electrons accelerate their relaxation and lead to the major involvement of the downshifted excitations in the generation of radio waves.
The Weibel component (with high oblique propagation, $k_\parallel << k_\perp$) of the EM excitations also becomes more apparent in the subsequent phases ($\omega_{pe}t > 160$).
We will return to it in a short comparative analysis in the next section.

In summary, our results are in agreement with previous studies \cite{Henri-etal-2019, Sauer-etal-2019, Lazar-etal-2023a}, demonstrating that the generation of radio emissions as escaping EM waves depends on the spectrum of the primary excitations, namely, their frequency and wavenumber, which in turn are conditioned by the properties of the electron beams.
Two modes can be responsible for the downshifted excitations with direct but also indirect involvement in radio emission processes.
In the early phase of our simulations, the $EB$ modes are excited and can play both roles. 
When two counterpropagating electron beams trigger pairwise, forward and backward propagating $EB$ excitations, these directly generate broad bands of the H component of radio waves, also downshifted from the (local) $2\omega_{pe}$.
Instead, the $EB$ excitations induced by a single (asymmetric) beam only contribute to the relaxation of the beam plasma distribution. 
The second mode responsible for downshifted excitations is represented by the $ZB$ waves induced later by the relaxed distributions. 
Radio waves can be generated by coupling waves of the same type $ZB$ in systems with two counterpropagating electron beams, or by the coalescence of $ZB$ and secondary waves (resulting from wave-wave decays) in systems with a single electron beam.

%%%%%%%%%%%%%%%%%%%%%%%%%%%%%%%%%%%%
\section{Conclusions and discussions}
%%%%%%%%%%%%%%%%%%%%%%%%%%%%%%%%%%%%  
Although downshifted ES waves have a well-established theoretical basis \cite{Cairns-1989, Gary-1993, Willes-Cairns-2000}, and have also been observed in association with electron beams in the Earth's foreshock \cite{Fuselier-etal-1985, Soucek-etal-2019}, their involvement in the context of solar radio emissions has not yet been addressed.
The most likely reason is that these excitations with lower frequencies, or downshifted with respect to the plasma frequency, do not conform to wave-wave interactions in the standard model of plasma radio emission \cite{Melrose_2008}.
Recent studies combining PIC simulations with rigorous predictions of wave dispersion and stability theory suggest, however, multiple implications of downshifted excitations in the generation of radio waves by electron beam plasmas \cite{Sauer-etal-2019, Lazar-etal-2023a}.
We have therefore refined PIC simulations that prove the indirect or even direct involvement of downshifted primary excitations in the generation of radio emission in electron beam plasmas.
Here we have analyzed in depth the primary excitations, while also accounting for the evolution of the electron beam at different time scales in radiative processes \cite{Kasaba-etal-2001, Thurgood-Tsiklauri-2015, Sauer-etal-2019}.
The main conclusions of our analysis can be sum up as follows.

When radio emissions are generated by the interaction of a single electron beam with the plasma, in particular, those satisfying condition~\eqref{e2}, the downshifted excitations are identified in two distinct stages.
For cases~1 and 2 (Table~\ref{t1}), the primary excitations from the early phases have frequencies close enough to the plasma frequency, the linear theory predicting ES instabilities of the electron-beam mode with frequency $\omega \lesssim \omega_{pe}$, still resonantly induced.
Simulations show that their spectra, although narrowed around the plasma frequency, still include both downshifted and upshifted frequencies, respectively, the electron-beam ($EB$) and $LB$ modes. 
Of which only the latter, i.e., $LB$ excitations, are consistent with the plasma model driving EM (daughter) radio waves.
However, the growing wave fluctuations act equally, whether they are upshifted or downshifted, on the electron beam, contributing to its plateau-on-tail or bump-on-tail relaxation.
Such electron distributions are observed in situ in association with Langmuir fluctuations and radio emission \cite{Ergun_1998}, and simulations confirm that they are still at the origin of the radio emission. 
The distributions with flattened beams, more like gentle bumps-on-tail, also result from the relaxation of the systems that satisfy condition~\eqref{e1} \cite{Kasaba-etal-2001,  Lee-etal-2019}.
Moreover, their persistence over time could also provide a plausible solution to Sturrock's dilemma.

As a consequence of this relaxation, the spectrum of primary excitations also evolves towards higher wave numbers, on the $LB$ branch with predominantly upshifted frequencies. 
Intense $LB$ excitations are specific to beams with sufficiently low density, as in case~1, when they lead to bursty radiative processes, both in ES and EM spectra, and both with fundamental (F) and harmonic (H) components. In such situations, the wave conversion can be linear, e.g., LMC type, as well as (weakly) nonlinear in resonant wave-wave interactions. 
However, the involvement of downshifted primary excitations remains unclear, especially because of the noisy and even bursty spectra.
%Indeed, the evolved plateau-on-tail distribution, similar to a gentle bump-on-tail, generates ES excitations with frequencies close to, but predominantly higher than the plasma frequency, with $\omega \gtrsim \omega_{pe}$, see our Figure~\ref{f1}. In this phase, the modes in the Langmuir branch are excited, and because this branch couples with that of the beam-electron modes, we can call them Langmuir-beam waves. These waves with $\omega \gtrsim \omega_{pe}$ are then able to satisfy the nonlinear decays from the standard model, e.g., $L \to L'+S$. Daughter waves are backward propagating Langmuir waves ($L'$)  of lower wavenumbers ($k_{L'} << k_L$), and low-frequency sound waves ($S$) satisfying $\omega_L = \omega_{L'} + \omega_S$ and $k_L = k_{L'} + k_S \simeq k_S$. 
%In the interval $360 < \omega_{pe} t < 500$, the spectra of these (non-linear) electrostatic excitations suggest a strong bursty discharge, both at the fundamental-F and higher harmonic-H frequencies, as well as at very low frequencies of the ion-acoustic or S-waves.
For denser and/or cooler beams, the parametric excitations resulting from the wave-wave decay of the $LB$ waves have lower intensities, but the spectra are less noisy, making it easier to distinguish the branch of the backscattered Langmuir ($L_-$) waves, the $L_+$ wave branch at very low wavenumbers (optical domain), and the low-frequency ion-sound waves $S_\pm$. 
Moreover, at larger time scales, the relaxed beams produce downshifted primary excitations, along the branch of the so-called $Z$-beam ($ZB$) mode.
These $ZB_+$ waves can couple with $L_-$ waves to explain the radio spectra obtained in this case, i.e., $ZB_+ + L_- \to 2T$, an H~$=2T$ emission more intense than the F component, also reported by the observations \cite{Bakunin-etal-1990, Reiner-MacDowall-2019}.
This mechanism appears to be sufficiently robust, as evidenced by the presence of the H component (albeit faint and obscured by other EM emission) until the end of our long-run simulations.
Thus, we infer for the first time that the downshifted excitations can have not only an indirect action in the initial phases of beam relaxation but also a subsequent direct involvement in the generation of escaping radio waves.

This direct contribution of the downshifted excitations to radio emissions can be highlighted much earlier, and can be even more significant, as it emerges from the analysis of case~3, with two counter-beams of electrons, symmetric and with the same properties as in case~2.
%Cases~2 and 3 are similar to those considered by \citeA{Ganse-etal-2012b}, who found indications of EM radio waves, both F and H emissions, only for the plasma system with two counterbeams.
In case~3, a symmetric spectrum of downshifted primary excitations is obtained, in both forward and backward propagation directions. 
In the early stages, the ES excitations are again around the plasma frequency and include both downshifted $EB$ and upshifted $LB$ modes. 
However, by contrast to case~2, the intense counter-propagating waves couple and generate, on much shorter time scales, a broadband of the H component of radio waves, highly isotropic and also downshifted with respect to $2 \omega_{pe}$.
It results from the involvement of the downshifted primary excitations (with frequency $\omega < \omega_{pe}$) in the coalescence processes, e.g., $EB_- + EB_+ \to 2T(\omega < 2\omega_{pe})$.
The energy conversion (from the kinetic energy of the electrons to ES and then EM waves) is significantly faster, and also contributes to the relaxation of the beams, reducing the time until only downshifted primary excitations are obtained.
These are $ZB$ excitations of lower intensity, this time symmetric and counterpropagating waves, which couple and generate the H radio emission, e.g., $ZB_+ + ZB_- \to 2T(\omega < 2\omega_{pe})$, still intense and highly isotropic. 

\begin{figure}[t]
   \centering   
   \includegraphics[width=\textwidth,trim={2cm 1.7cm 2.cm 1.5cm},clip]{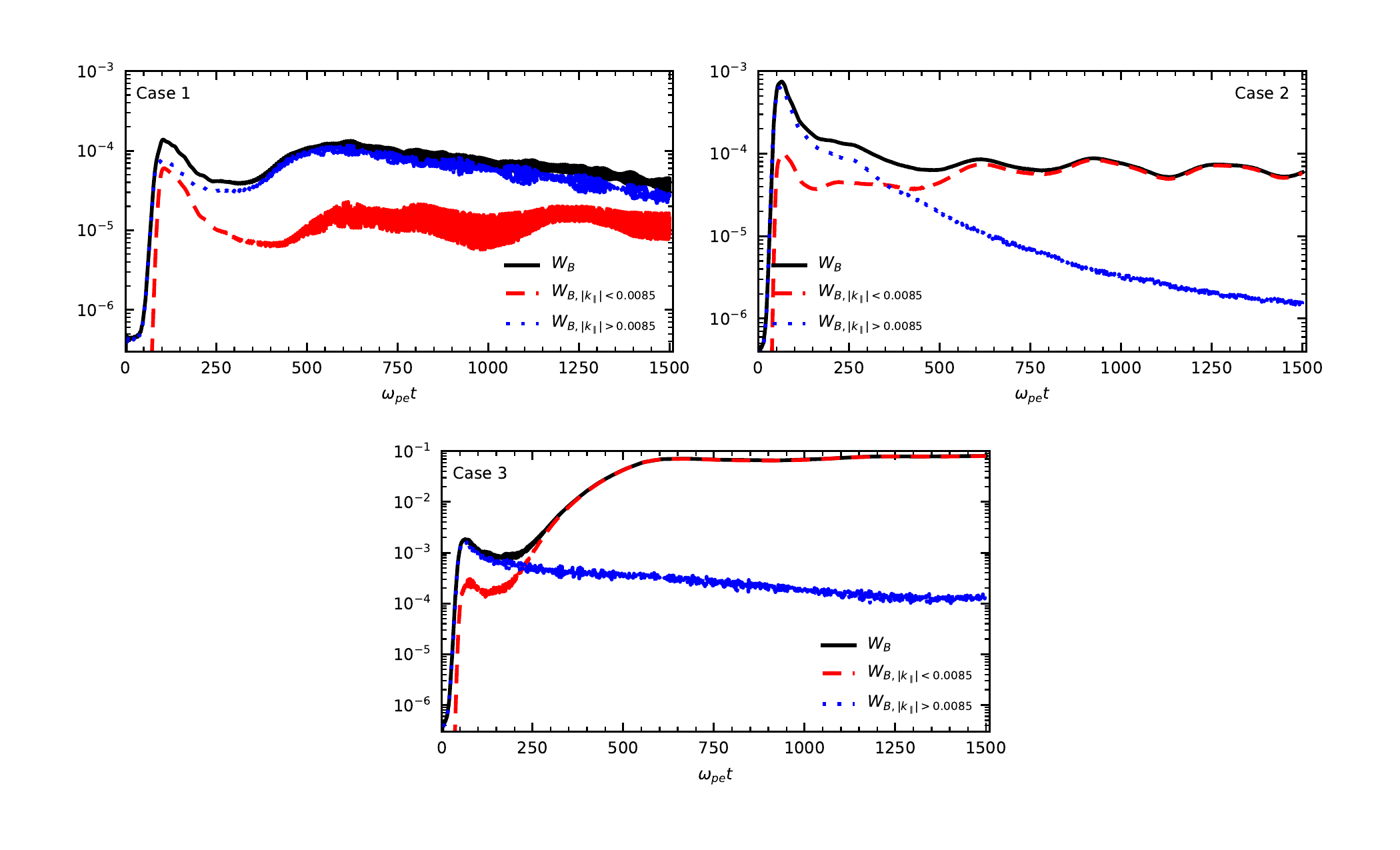}
\begin{internallinenumbers}
   \caption{\small Partition of the wave magnetic energy density $W_B$ (black), normalized to the initial kinetic energy of the beam, for cases~1, 2, and 3, between the highly oblique Weibel-like instability (dashed red) and the EM radio emission (blue dotted).}
   \label{f10}
\end{internallinenumbers}
   \end{figure}

The radiative mechanisms triggered by two counterbeams seem to be the most robust, but we should not ignore the Weibel (or filamentation) instabilities, which are also primary excitations, but of EM nature, i.e., the so-called EM electron beam modes. 
\citeA{Karlicky-2009} and \citeA{Thurgood-Tsiklauri-2015} pointed out the competitiveness of these excitations, by taking over a significant part of the EM wave energy. 
For the same parameterization as in case 2 but in the absence of the uniform magnetic field (unmagnetized plasma), the simulations of \citeA{Thurgood-Tsiklauri-2015} did not lead to radio emissions, while \citeA{Lazar-etal-2023a} already found that the presence of the magnetic field has a favorable influence on the production of radio emissions. 
Our refined simulations (with the same $\omega_{pe}/ |\Omega_e| =100$) confirm this effect even for such dense electron beams. 
However, the background magnetic field does not have a direct influence on the ES excitations, but most likely, it inhibits other competing EM excitations, such as Weibel-like EM instabilities of the highly-oblique (perpendicular) O-mode \cite{Karlicky-2009, Lazar-etal-2010}.
Fig.~\ref{f10} shows the total fluctuating magnetic energy density $W_B= \int dx dy \delta B_z^2 / K_{b,0}$ (black), normalized to the initial kinetic energy density of the beam, $K_{b,0}$, for all our runs, case~1 (upper), 2 (middle) and 3 (lower).
$W_B$ is partitioned between the EM radio emission (blue dotted) and the perpendicular propagating mode (red dashed), most probably a Weibel-like excitation.
In the initial phase ($\omega_{pe}t < 200$), the best conversion of the kinetic energy of the electrons into EM radio emissions occurs in case~3 with a little over 0.1~\%; similarly, in case~2 with an efficiency slightly below 0.1~\%, while in case 1, the efficiency is not much below 0.01~\% and only slightly in favor of radio emissions.
The radio spectra in case~2 have a peculiarity regarding the contrast between the F and H emissions, the latter being more intense. 
Moreover, this contrast increases in case~3 with two counterbeams of electrons, when the H emission becomes much more intense and the F component is almost absent. 
These characteristics can help identify certain radio plasma sources, in particular those with counterbeams of electrons with similar properties, i.e. (almost) symmetric counterbeams as those considered in case~3.

At larger time scales in Figure~\ref{f10}, the energy conversion remains favorable to radio emissions only in case~1, i.e, the blue-dotted curve remains always above the red-dashed, while in the other two cases the (relative) levels of radio emissions decrease, much steeper in case~2. 
However, in cases~2 and 3, the late efficiency of the Weibel excitations becomes very high, in case 3 reaching to convert almost 10~\% of the initial kinetic energy of the electron beam.
This is justified by the relaxation of the electron beam plasma to an anisotropic distribution with an effective temperature (or thermal spread) anisotropy still favorable to Weibel-type instabilities, but not to ES excitations at the origin of radio emissions.
%Beyond the plateau of $W_B \simeq$ constant, the conversion to the EM radio emission decreases, while the increase of $W_B$ becomes mainly due to the Weibel-like mode.
We can therefore expect that for weakly magnetized plasma systems, such as those considered here and specific to large heliocentric distances, e.g., at 1~AU and beyond, the spectra of radio emissions (especially the F component, see cases~1 and 2)
may be obscured by Weibel-like excitations. 
Instead, the production of radio emissions, as well as the direct role of downshifted excitations can be significantly enhanced, both in the early and late phases, if the magnetic fields are more intense, as, for instance, in sources of type II emission linked to shocks produced closer to the Sun, or in plasma sources of type III bursts from coronal flares.
Recent PIC simulations confirm that in sufficiently strong magnetic fields the electron beam plasma interaction leads to an intense excitation in the field-aligned longitudinal mode, and to a significantly enhanced H emission, though the underlying wave-wave interactions in the magnetoactive plasma still remain to be elucidated \cite{Lee-etal-2022}.

To conclude, our analysis provides new arguments for the existence of extended regimes of electron beam plasmas (less debated so far), capable of producing radio emissions such as type II and type III solar radio bursts.
%Such in-depth analyzes have multiple merits, starting with the understanding of the physical mechanisms at the origin of radio emissions, and ending with their applications, such as the remote diagnosis of their plasma sources.
Further investigation requires the relaxation of electron beams, in particular, the relaxed plateau-on-distributions that are still found to be effective in generating radio emissions. 
These radiative regimes can even involve reduced excitations, while their persistence in time has a particular relevance in solving Sturrock's dilemma.
Such radio spectra dominated eventually by a (second) H radio emission have extended relevance in type III solar bursts \cite{Reiner-MacDowall-2019, Jebaraj_2023}.
Another important conclusion is that the electron beams at the origin of the radio emissions do not necessarily have to have very low densities, as required in the standard plasma model for the excitation of primary Langmuir waves.
Therefore, the present results motivate the upgrades of the standard model of radio plasma emission to include the new extended regimes of electron beams and downshifted primary excitations. 
This also has important implications in the remote diagnosis of radio plasma sources which, in general, is based on such a standard model.

%%%%%%%%%%%%%%%%%%%%%%%%%%%%%%%%%%%%%%%%%%%%%%%
%% Optional Appendices go here
%
\appendix %First online appendix

%%%%%%%%%%%%%%%%%%%%%%%%%%%%%%%%%%%%%
%\section{Insights from linear theory} \label{linear-theory}
%%%%%%%%%%%%%%%%%%%%%%%%%%%%%%%%%%%%%

%Instabilities of the Langmuir waves are shown in Fig.~\ref{f3} with, e.g., green and blue solid lines in left panels.
%

%%%%%%%%%%%%%%%%%%%%%%%%%%%%%%
%\section{Non-filtered spectra}

%%%%%%%%%%%%%%%%%%%%%%%%%%%%%%%%%%%%%%%%%%%
% DATA SECTION and ACKNOWLEDGMENTS
%%%%%%%%%%%%%%%%%%%%%%%%%%%%%%%%%%%%%%%%%%

\section*{Open Research Section}
The simulation code we have used is adapted from the publicly available KEMPO1 code from \citeA{Matsumoto1993}. The linear solutions used were obtained with the DIS-K code available at \url{https://github.com/ralopezh/dis-k} \cite{Lopez-etal-2021, Lopez2023}. The plots were produced with Matplotlib, available under the Matplotlib license at \url{https://matplotlib.org/}. The relevant input data are available at \citeA{Lopez2024}.

\acknowledgments
The authors acknowledge support from the Ruhr-University Bochum, the Katholieke Universiteit Leuven, and Qatar University. These results were also obtained in the framework of the projects C14/19/089 (C1 project Internal Funds KU Leuven), G0B5823N (FWO-Vlaanderen), WEAVE project G002523N / FI~706/31-1 (FWO-Vlaanderen / DFG-Germany), 4000134474 (SIDC Data Exploitation, ESA Prodex), Belspo project B2/191/P1/SWiM. Powered@NLHPC: This research was partially supported by the supercomputing infrastructure of the NLHPC (CCSS210001). The resources and services used in this work were provided by the VSC (Flemish Supercomputer Center), funded by the Research Foundation - Flanders (FWO) and the Flemish Government. The open access publication of this article was funded by Qatar National Library.

%%%%%%%%%%%%%%%%%%%%%%%%%%%%%%%%%%%%%%%%%%%
% REFERENCES and BIBLIOGRAPHY
%%%%%%%%%%%%%%%%%%%%%%%%%%%%%%%%%%%%%%%%%%

\bibliography{ rad-dsw}

% Please use ONLY \cite and \citeA for reference citations.
% \cite for parenthetical references
% ...as shown in recent studies (Simpson et al., 2019)
% \citeA for in-text citations
% ...Simpson et al. (2019) have shown...

\end{document}